\documentclass[iop]{emulateapj-rtx4}
\usepackage{bm}
\usepackage{threeparttable}
\usepackage{CJK}
\newcommand{\alphasc}{\ensuremath{\alpha_{\mathrm{sc}}}} 

\newcommand{\alphaMLT}{\ensuremath{\alpha_{\mathrm{MLT}}}}	
\newcommand{\alphaOv}{\ensuremath{\alpha_{\mathrm{ov}}}}	

\def\andname{ }

\def\bm{}

\def\bfnabla{\ensuremath{\nabla}}

\def\msun{M_\odot}
\def\lsun{L_\odot}
\def\rsun{R_\odot}

\renewcommand\bv{\boldsymbol{ v}}

\newcommand\bF{\boldsymbol{ F}}
\newcommand\bP{\boldsymbol{ P}}
\newcommand\bn{\boldsymbol{ n}}

\newcommand\bfr{{\sf\boldsymbol{ f}}}

\newcommand{\kb}{\mathrm{k_B}}

\def\<{\,\langle\langle}
\def\>{\,\rangle\rangle}

\shorttitle{Radiation Hydro Simulations of Massive Star Envelopes}

\begin{document}
\begin{CJK*}{UTF8}{gbsn}

\shortauthors{Y.-F. Jiang et al.}
\author{Yan-Fei Jiang (姜燕飞)\altaffilmark{1\footnote{Einstein Fellow}}, Matteo Cantiello\altaffilmark{2}, Lars Bildsten\altaffilmark{2,4}, Eliot Quataert\altaffilmark{3}, 
Omer Blaes\altaffilmark{4}}
\affil{$^1$Harvard-Smithsonian Center for Astrophysics, 60 Garden Street, Cambridge, MA 02138, USA} 
\affil{$^2$Kavli Institute for Theoretical Physics, University of California, Santa Barbara, CA 93106, USA} 
\affil{$^3$Astronomy Department and Theoretical Astrophysics Center, 
University of California at Berkeley, Berkeley, CA 94720-3411, USA} 
\affil{$^4$Department of Physics, University of California, Santa Barbara, CA 93106, USA}

\title{Local Radiation Hydrodynamic Simulations of Massive Star Envelopes at the Iron Opacity Peak}

\begin{abstract}

  We perform three dimensional radiation hydrodynamic simulations of the
  structure and dynamics of radiation dominated envelopes of massive
  stars at the location of the iron opacity peak. One dimensional
  hydrostatic calculations predict an unstable density inversion at
  this location, whereas our simulations reveal a complex interplay of
  convective and radiative transport whose behavior depends on the
  ratio of the photon diffusion time to the dynamical time. The latter
  is set by the ratio of the optical depth per pressure scale height,
  $\tau_0$, to $\tau_c=c/c_g$, where $c_g\approx 50\ {\rm km \
  s^{-1}}$ is the isothermal sound speed in the gas alone.  When
  $\tau_0\gg\tau_c$, convection reduces the radiation acceleration and
  removes the density inversion. The turbulent energy transport in the
  simulations agrees with mixing length theory and provides its first
  numerical calibration in the radiation dominated regime.  When
  $\tau_0 \ll \tau_c$, convection becomes inefficient and the
  turbulent energy transport is negligible. The turbulent velocities
  exceed $c_g$, driving shocks and large density fluctuations that
  allow photons to preferentially diffuse out through low-density
  regions. However, the effective radiation acceleration is still
  larger than the gravitational acceleration so that the time average
  density profile contains a modest density inversion. In addition,
  the simulated envelope undergoes large-scale oscillations with
  periods of a few hours. The turbulent velocity field may affect the
  broadening of spectral lines and therefore stellar rotation
  measurements in massive stars, while the time variable outer
  atmosphere could lead to variations in their mass loss and stellar
  radius.
  

\end{abstract}

\maketitle

\section{Introduction}
Massive stars play an important role in many astrophysical
environments. The radiative and mechanical energy output from massive
stars are important feedback mechanisms regulating star formation and
the structure of the interstellar medium in galaxies
\citep[][]{Kennicutt2005,Smith2014}.  Radiation from massive stars
also plays an important role in the reionization of the universe
\citep[][]{BrommLarson2004}.  When massive stars explode, they produce
various types of supernovae and high energy transients, while leaving
behind black holes and neutron stars whose properties depend on the
previous history of massive stellar evolution
\citep[][]{Hegeretal2003}.

Because of the steep mass-luminosity relation of main sequence stars,
the luminosity approaches the Eddington-limit for electron scattering
in stars with masses larger than $\sim 50-100\msun$
\citep[][]{Crowther2007,Maederetal2012,Sanyaletal2015}.  In massive
star envelopes with non-zero metallicity, the opacity
increases significantly compared with electron scattering due to iron
opacity at temperatures $\approx  1.8\times 10^5$ K (e.g.,
\citealt{Paxtonetal2013} and references therein). As a result, the
actual radiation force can locally exceed the gravitational force near
the iron opacity peak.  This same physics can operate at somewhat
lower temperatures due to the opacity produced by helium and
hydrogen \citep[e.g.][]{Maeder1981}. 

The locally super-Eddington luminosity near the iron opacity peak (and
related opacity peaks) has been a long standing challenge for one
dimensional (1D) stellar evolution models of massive
stars. Hydrostatic and thermal equilibrium models with a
super-Eddington luminosity require density and gas pressure inversions
\citep[][]{Joss:1973,Graefener2012,Paxtonetal2013,Owocki2015}, which are usually convectively
unstable.  When the iron opacity peak is sufficiently deep in the
star, convection is efficient and can rearrange the stellar structure
to have constant entropy, removing the density inversion predicted by
radiative equilibrium models.  When the iron opacity peak is close to
the surface, however, convection is inefficient and cannot carry the
necessary super-Eddington flux. 
It is unclear how the stellar structure adjusts in
this limit, and in particular if the density inversions are stable \citep{Maeder2009,Sanyaletal2015}.
This results in very uncertain stellar radii in the very massive star
regime \citep{Yusof2013,Koheler2015}. 
To complicate the situation even further, stellar evolution calculations of massive stars become numerically difficult when the opacity near the stellar
photosphere significantly exceeds the electron scattering opacity. 
In this situation  the convective energy transport is often artificially enhanced in 1D
stellar evolution models \citep[][]{Maeder1987,Paxtonetal2013}, which leads to the disappearance of the density inversions \citep{Stothers1979}.

The density and temperature dependence of the opacity near
composition-specific opacity peaks can also drive radiation
hydrodynamic instabilities distinct from convection
\citep[][]{Kiriakidis1993,BlaesSocrates2003,Suarezetal2013}.  In many
cases, these cannot be captured by 1D stellar evolution models.  These
instabilities have been suggested as the origin of the large mass loss
rates from massive stars that cannot be explained by line-driven winds
\citep[][]{Glatzel1993}.  The amount of mass loss by line driven winds
\citep[][]{Pulsetal2008} also depends on the structure of the stellar
photosphere (e.g., the amplitude of density and velocity fluctuations, potentially affecting the degree of wind clumping).

In this paper, we present three-dimensional (3D) radiation
hydrodynamic simulations of stellar envelopes near the iron opacity
peak and quantify the impact of convective and radiation hydrodynamic
instabilities on the stellar envelopes' properties and dynamics.
These calculations provide the first direct calibration of classical
mixing length theory (MLT) \citep[][]{Cox1968} in the radiation
pressure dominated regime.
In addition, we quantify the extent to which convection and/or other
hydrodynamic instabilities generate a ``porous'' atmosphere that
substantially increases the radiation flux relative to that predicted
by one-dimensional models.  This is critical for determining whether
locally super-Eddington regions in a stellar envelope can initiate a
large-scale stellar wind via continuum radiation pressure or whether
the stellar radiation freely escapes through low density channels
\citep[][]{Shaviv1998,vanMarle2008}.

Our work utilizes numerical techniques for accurate 3D radiation
hydrodynamic simulations based on a variable Eddington tensor (VET)
formalism \citep[][]{Jiangetal2012,Davisetal2012}.  These techniques
have been successfully used to study a wide range of astrophysical
problems \citep[][]{Jiangetal2013a,Jiangetal2013c,Jiangetal2014} and
have significant advantages relative to ad-hoc closures such as the
flux-limited diffusion (FLD) approximation
\citep[][]{Davisetal2014,Progaetal2014}.  The direct solution of the
radiation transfer equation (via VET or Monte-Carlo methods) is
particularly important for studying mildly optically thick regions
near the stellar photosphere.  Indeed, recent studies have shown that
approximate radiation transfer schemes such as FLD or M1 can reach
completely opposite conclusions compared with VET or Monte Carlo
methods \citep[][]{Davisetal2014,TsangMilosavljevic2015}.

The remainder of this paper is organized as follows.  In \S
\ref{sec:mesamodels} we identify the key dimensionless parameter that
governs the physics of near surface convection zones in stellar
envelopes.  We also describe the 1D stellar models that motivate the
specific conditions in our 3D radiation hydrodynamic simulations.  In
\S \ref{sec:method} we describe our numerical methods while in \S
\ref{sec:results} we present our results.  Finally, in \S
\ref{sec:conc} we summarize our key results and discuss their
implications.

\section{The Model Parameters}
\label{sec:mesamodels}
The relative importance of convective and radiative energy transport
can be characterized by a critical optical depth $\tau_c$ over the
pressure scale height $H_0$ near the iron opacity peak. As we will
show based on our simulations, for conditions typical of massive
stars, the maximum turbulent velocity reaches the isothermal sound
speed when convection starts to become inefficient.  Then the critical
optical depth can be estimated by balancing the diffusive radiation
flux $\sim a_rT_0^4c/\tau$ and the maximal convective flux in the
radiation pressure dominated regime $\sim c_{g,0}a_rT_0^4$:
\begin{eqnarray}
\tau_c=c/c_{g,0}.
\label{eq:tauc}
\end{eqnarray}
We define the gas isothermal sound speed $c_{g,0}$, radiation sound
speed $c_{s,0}$, pressure scale height $H_0$, optical depth $\tau_0$
and typical dynamic time scale $t_0$ based on the density $\rho_0$ and
temperature $T_0$ at the iron opacity peak
\begin{eqnarray}
c_{g,0}&=&\sqrt{\frac{P_0}{\rho_0}},\nonumber\\
c_{s,0}&=&\sqrt{\frac{a_rT_0^4}{3\rho_0}},\nonumber\\
H_0&=&\frac{c_{s,0}^2}{g}=\frac{a_rT_0^4}{3\rho_0g},\nonumber\\
\tau_0&=&\kappa_t(\rho_0,T_0)\rho_0H_0,\nonumber\\
t_0&=&\frac{H_0}{c_{s,0}}.
\end{eqnarray}
The characteristic radiation pressure is $P_{r,0}=a_rT_0^4/3$ while
the gas pressure is $P_{0}$.

\begin{figure}[htp]
\begin{center}
\includegraphics[width=1.0\columnwidth]{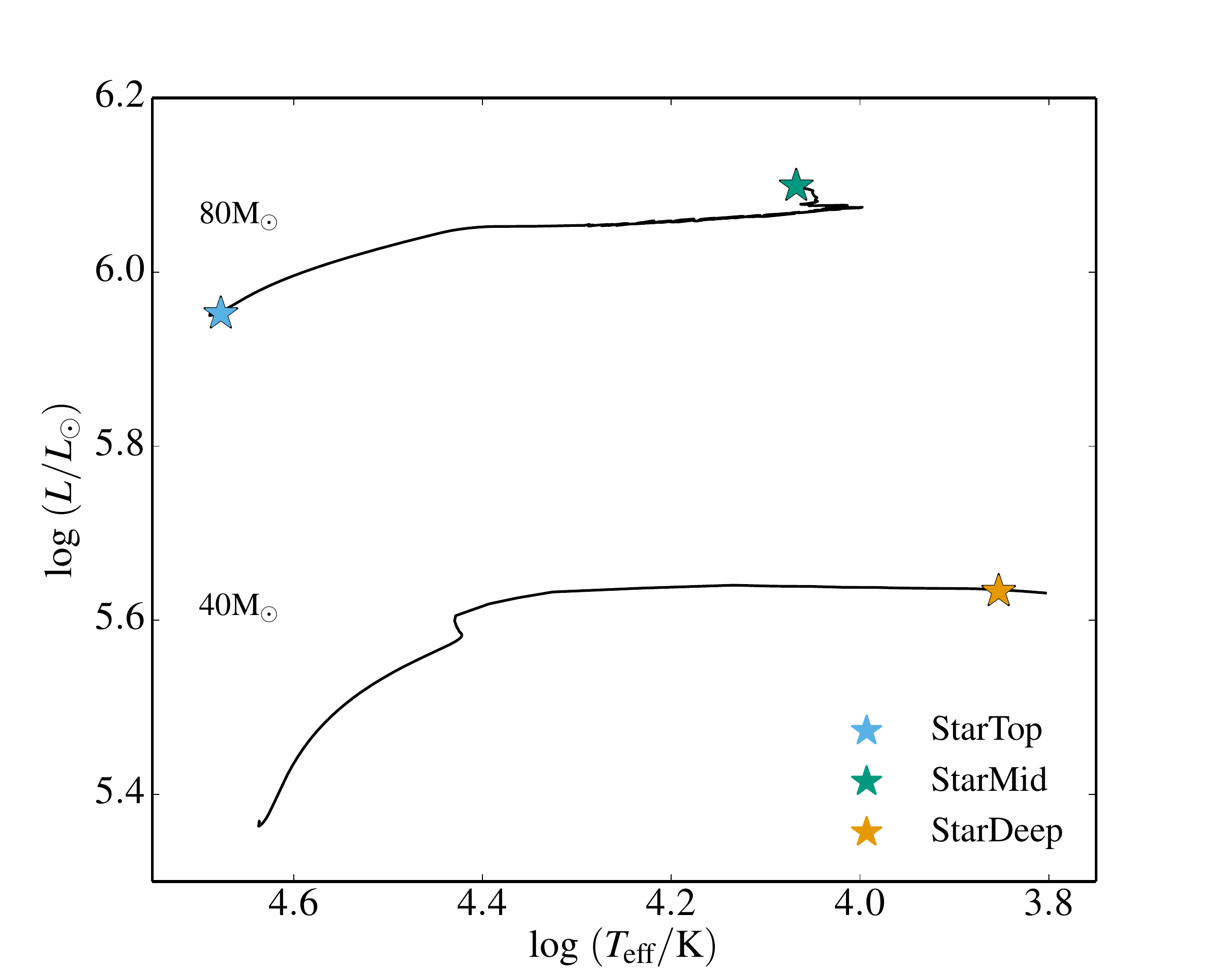}
\caption{Stellar evolution tracks of two solar metallicity stars with initial mass of $40\msun$ 
and $80\msun$ calculated with the 1D stellar evolution code MESA. 
The stellar symbols identify the locations corresponding to the three initial conditions of the 
3D radiation hydro calculations with {\sc Athena} listed in Table \ref{Table:parameters}. 
The {\sc Athena} calculations are effectively local plane-parallel atmosphere calculations 
motivated by the MESA models.
\label{HRD}%
}
\end{center}
\end{figure}

A related way to interpret the critical optical depth $\tau_c$ in
equation (\ref{eq:tauc}) is that the typical thermal time scale due to
photon diffusion $t_c$ near the iron opacity peak is given by
\begin{eqnarray}
t_c=\frac{\tau_0H_0}{c}=\frac{\tau_0}{\tau_c}\frac{c_{s,0}}{c_{g,0}}t_0. 
\label{eq:tc}
\end{eqnarray}
When $\tau_0\gg\tau_c$, the diffusion time scale is longer than the
local dynamic time scale and we expect convection to be an efficient
energy transport mechanism.  When $\tau_0\ll\tau_c$, however,
radiative diffusion is rapid enough that convection may behave very
differently compared with classical mixing length theory.  
This can also be related to the P\'eclet number $P_e$, which is the ratio of advective 
to diffusive flux. When $\tau_0\gg \tau_c$, we expect $P_e\gtrsim 1$, while when 
$\tau_0\ll\tau_c$, $P_e\lesssim 1$. 
Note that there is some ambiguity as to whether $\tau_c$ should be defined in
terms of the gas isothermal sound speed $c_{g,0}$ or the radiation
sound speed $c_{s,0}$.  As we shall show, we find that the former
better characterizes the maximum velocities reached in our
simulations.

In order to identify stellar conditions spanning the different regimes
of radiation dominated convection, we use Modules for Experiments in
Stellar Evolution (MESA, \citet{Paxtonetal2011,Paxtonetal2013,Paxton:2015}
release r7385) to evolve stars with initial mass $40\msun$ and
$80\msun$.  The models have an initial metallicity $Z=0.02$ with a
mixture taken from \citet{Asplund:2005} and are calculated using the
OPAL opacity tables \citep{iglesias96}.  Convection is accounted for
using mixing-length theory (MLT) in the \citet{Henyey:1965}
formulation with $\alphaMLT=1.5$.  Convective boundaries are
determined using the Ledoux criterion and we use semiconvection with
an efficiency $\alphasc$ = 0.01 \citep{Langer:1983,Langer:1985}. An
exponentially decaying overshooting \citep{Herwig:2000} has been
included with $\alphaOv=0.001$ above and below convective
boundaries. Stellar winds are included according to the mass loss
scheme described in \citet{Glebbeek:2009} (Dutch wind scheme) with an
efficiency parameter $\eta = 0.8$.
 
We have identified specific times in the MESA massive star models with
conditions that range from $\tau_0/\tau_c \ll 1$ to
$\tau_0/\tau_c \gg 1$.  We label these models as {\sf StarDeep}, {\sf
  StarMid} and {\sf StarTop}; they have $\tau_0/\tau_c=15.2,0.97$ and
$0.025$, respectively.  Figure \ref{HRD} shows the positions of these
models in the HR diagram.  Model {\sf StarDeep} has an initial mass of
$40\msun$. It is evolved to an age of $4.3\times10^6$ year in MESA at
which point it has an effective temperature $7.13\times10^3$ K, radius
$431\rsun$ and luminosity $8.96\times 10^5\lsun$.  Model {\sf StarMid}
starts from a $80\msun$ star.  At an age of $3.02\times 10^6$ year it
has an effective temperature $1.17\times 10^4$ K, radius $274\rsun$
and luminosity $1.26\times10^6\lsun$.  Model {\sf StarTop} also starts
from an $80\msun$ star but is only evolved to an age of $9.15\times10^4$
year, at which point it has an effective temperature $4.75\times10^4$
K, radius $14\rsun$ and luminosity $8.96\times10^5\lsun$.

As described in \S \ref{sec:method}, we use plane-parallel 3D
radiation hydrodynamics simulations to study the physics of massive
stellar envelopes.  The initial radial location of the iron opacity
peak $r_0$ and the density $\rho_0$, temperature $T_0$ and luminosity
$L_0$ at this location are taken from the corresponding MESA models.
The fiducial parameters of our
initial conditions are listed in Table \ref{Table:parameters}.


\section{Numerical Method}
\label{sec:method}
\subsection{Equations}
We model the envelopes of massive stars in plane parallel geometry,
neglecting the global stellar geometry. This is suitable for studying
the effects of 3D radiation hydrodynamics on energy transport in the
stellar envelope, but is not appropriate for studying global effects
such as stellar outflows.  We solve the frequency-independent (gray)
radiation hydrodynamic equations in cartesian coordinates $(x,y,z)$
with unit vectors ($\bm{\hat{x}}$, $\bm{\hat{y}}$, $\bm{\hat{z}}$).
The gravitational acceleration $g$ is assumed to be a constant and
along the $-\bm{\hat{z}}$ direction. As the mass of the envelope is
less than $1\%$ of the core mass, this is a reasonable assumption (the
slight change in $1/r^2$ in the stellar envelope is also neglected).
The equations are \citep[e.g.,][]{Jiangetal2012,Jiangetal2013a}
\begin{eqnarray}\label{eqn:equations}
\frac{\partial\rho}{\partial t}+\bfnabla\cdot(\rho \bv)&=&0, \nonumber \\
\frac{\partial( \rho\bv)}{\partial t}+\bfnabla\cdot({\rho \bv\bv+P{{\sf I}}}) &=&-\bm{ S_r}(\bP)- \rho g\bm{ \hat z},\  \nonumber \\
\frac{\partial{E}}{\partial t}+\bfnabla\cdot\left[(E+P)\bv\right]&=&-cS_r(E)-g\rho\bv\cdot\bm{\hat z},  \nonumber \\
\frac{\partial E_r}{\partial t}+\bfnabla\cdot \bF_r&=&cS_r(E), \nonumber \\
\frac{1}{c^2}\frac{\partial \bF_r}{\partial t}+\bfnabla\cdot{\sf P}_r&=&{\bf S_r}(\bP),
\end{eqnarray}
where the radiation source terms are
\begin{eqnarray}\label{eqn:sources}
{\bf S_r}(\bP)&=&-\rho\left(\kappa_{aF}+\kappa_{sF}\right)\left[\bF_r-\left(\bv E_r+\bv\cdot{\sf P}_r\right)\right]/c \nonumber \\
&+&\rho\bv(\kappa_{aP}a_rT^4-\kappa_{aE}E_r)/c,\nonumber\\
S_r(E)&=&\rho(\kappa_{aP}a_rT^4-\kappa_{aE}E_r) \nonumber \\
&+&\rho(\kappa_{aF}-\kappa_{sF})\frac{\bv}{c^2}\cdot\left[\bF_r-\left(\bv E_r+\bv\cdot{\sf  P}_r\right)\right].
\end{eqnarray}
Notice that local thermal equilibrium (LTE) is assumed for the emission term. 

In the above equations, $\rho,P,\bv,c$ are the gas density, pressure,
flow velocity and speed of light respectively.  The total gas energy
density is $E=E_g+\rho v^2/2$, where $E_g=P/(\gamma-1)$ is the
internal gas energy density with a constant adiabatic index
$\gamma=5/3$.  The gas pressure is $P=\rho \kb T/\mu$, where
$\kb$ is Boltzmann's constant and $\mu=0.62m_p$
is the mean molecular weight for nearly fully ionized gas with proton 
mass $m_p$.  The
radiation constant is $a_r=7.57\times10^{15}$ erg cm$^{-3}$ K$^{-4}$,
while $E_r, \bF_r$ are the radiation energy density and flux.  The Rosseland
 mean absorption and scattering opacities are denoted by 
$\kappa_{aF}$ and $\kappa_{sF}$, while $\kappa_{aP}$ and $\kappa_{aE}$ are the Planck
and energy mean absorption opacities.  We describe our choice of these
opacities in the following section.  

Equations (\ref{eqn:equations}) and (\ref{eqn:sources}) are solved in
the mixed frame, meaning that the radiation flux $\bF_r$ and energy
density $E_r$ are Eulerian variables, while the material-radiation
interaction terms in equations (\ref{eqn:sources}) are written in the
co-moving frame \citep[e.g.,][]{Lowrieetal1999}.  The Eulerian and
co-moving flux $\bF_{r,0}$ are related through
$\bF_{r,0}=\bF _r-\left(\bv E_r+\bv\cdot{\sf P} _r\right)$. Notice
that only the co-moving radiation flux contributes to the radiation
acceleration while the advection flux does not. 

We relate the radiation pressure ${\sf P}_r$ and radiation energy
density through a variable Eddington tensor (VET),
${\sf P}_r=\bfr E_r$.  We do not use an assumed closure relation (such
as FLD or M1) to specify $\bfr$. Instead, we compute it directly from
angular quadratures of the specific intensity $I_r$ for each angle
$\bn$, which is calculated from the time-independent radiation
transfer equation
\begin{eqnarray}
\frac{\partial I_r}{\partial s}=\kappa_t\left(S-I_r\right),
\label{eqn:vetequation}
\end{eqnarray}
where $S$ is the radiation source term and $\kappa_t$ is the total
specific opacity.  Because the typical dynamic time scale
is much longer than the light crossing time, $\bfr$ barely changes
between each time step. Our use of the time
independent transfer equation to calculate the angular distribution of
the radiation field is thus very accurate.  We solve the transfer
equation in equation (\ref{eqn:vetequation}) using the method of short
characteristics, as described in \cite{Davisetal2012}. Then $\bfr$ is
calculated according to its definition
\begin{eqnarray}
\bfr=\frac{\int I_r\bn\bn d\Omega}{\int I_rd\Omega},
\label{eq:eddten}
\end{eqnarray}
where $\Omega$ is the solid angle. We emphasize that we only 
use the time-independent transfer equation (\ref{eqn:vetequation}) 
to calculate the Eddington tensor $\bfr$, which represents the angular 
distribution of the radiation field. The moments of $I_r$ are not used 
to calculate the radiation-material interactions. Instead, the radiation moments 
$E_r$, $\bF_r$ and the radiation source terms are calculated self-consistently 
through the time dependent radiation moment equations (\ref{eqn:equations}).

The Eddington tensor in equation (\ref{eq:eddten}) closes the radiation
moment equations (eqs. \ref{eqn:equations}).  We solve these equations
using the Godunov radiation MHD code {\sc Athena}, as described and
tested in \cite{Jiangetal2012} (and with additional improvements
described in \cite{Jiangetal2013b}).  We solve the radiation
subsystems as well as the radiation source terms implicitly, while the
hydro equations are solved explicitly as in the original {\sc Athena}
code \citep[][]{Stoneetal2008}, with appropriate modifications due to
the stiff radiation source terms \citep[][]{Jiangetal2012}.

\subsection{The Opacity}
\label{sec:opacity} 

In order to correctly capture the temperature and density dependencies
of the opacity, particularly near the iron opacity peak, we directly
use the opacity table from MESA
\citep[][]{Paxtonetal2011,Paxtonetal2013, Paxton:2015} in our {\sc
  Athena} simulations.  We assume a metallicity $Z=0.02$ and hydrogen
fraction $X=0.6$.  For the density $\rho$ and temperature $T$ in each
cell, the total specific opacity $\kappa_t$ is calculated with
bilinear interpolation based on the table.  This is used as the total
specific flux mean opacity $\kappa_t=\kappa_{aF}+\kappa_{sF}$.
In order to split this into scattering and absorption opacities, we
assume a constant electron scattering opacity $\kappa_{sF}=0.32$
g$^{-1}$ cm$^{2}$.  All of the absorption opacities
($\kappa_{aF},\kappa_{aP},\kappa_{aE}$) are assumed to be the same for
simplicity.

\begin{figure}[h!]
\begin{center}
\includegraphics[width=1.0\columnwidth]{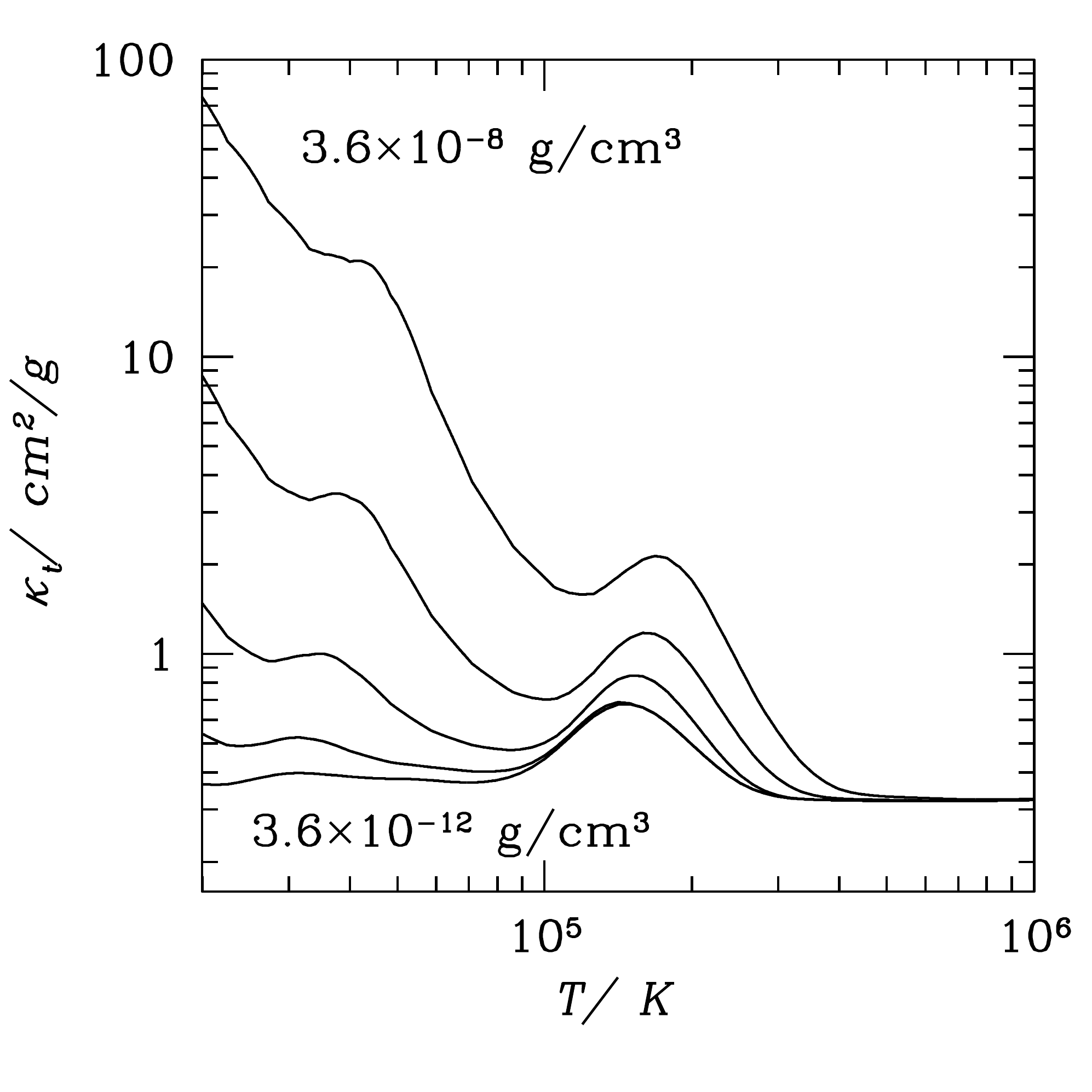}
\caption{The opacity $\kappa_t$ as a function of density and
temperature near the iron opacity peak at $T\approx 1.5\times 10^{5}$ K.
Each line shows the opacity as a function of temperature for a fixed density. From 
top to bottom, the density decreases by a factor of $10$ between
neighboring lines. 
\label{opacitytable}%
}
\end{center}
\end{figure}

Figure \ref {opacitytable} shows the actual opacity $\kappa_t$ we use
for a range of temperatures and densities appropriate to stellar
envelopes. The opacity peak between $10^5$ and $2\times10^5$ K is due
to Fe, which is the region we focus on.  The increase in opacity
at lower temperatures is due to He and H.  The iron opacity peak has a
sensitive dependence on temperature but a weak dependence on density.
Particularly when $\rho\lesssim 3.6\times10^{-11}$ g cm$^{-3}$,
$\kappa_t$ does not depend on density for $T \gtrsim 10^5$ K.  For the
stellar envelopes we study, the temperature usually does not drop
below $4\times10^4$ K.  Moreover, when this does happen, the density
is always smaller than $10^{-10}$ g/cm$^{3}$.  As a result, the He and
H opacity peaks usually do not show up in our calculations.

\subsection{Initial and Boundary Conditions}
\label{sec:inibd}

We set our initial conditions in the {\sc Athena} simulations by using
the MESA models described in \S \ref{sec:mesamodels} to determine the
initial radial location $r_0$, density $\rho_0$ and temperature $T_0$
at the iron opacity peak.  The gravitational acceleration $g$ is also
chosen to be the value at $r_0$ in the MESA models. We then apply a
constant radiation flux $F_{r,i}$ through the whole simulation box,
which represents the radiation coming from the center of the massive
star. The initial radiation flux is determined by the total
luminosity $L_0$ in the MESA models at $r_0$:
$F_{r,i}=L_0/(4\pi r_0^2)$.

With these fiducial parameters and the opacity given in \S
\ref{sec:opacity}, we construct our initial conditions by solving for
a hydrostatic envelope in thermal equilibrium.  Starting from $z=r_0$,
all the fluid and radiation quantities are integrated as a function of
height using
\begin{eqnarray}
\frac{\partial P}{\partial z}&=&\frac{\rho\kappa_t}{c} F_{r,i}-\rho g,\nonumber \\
E_r&=&a_rT^4,\nonumber\\
\frac{1}{3}\frac{\partial E_r}{\partial z}&=&-\frac{\rho\kappa_t}{c} F_{r,i}.
\end{eqnarray}
We integrate above and below $r_0$ to capture the whole profile of the
iron opacity peak region in our simulation box.  We then add random
perturbations with $0.01\%$ amplitude to the density in the simulation
box of dimensions $L_x,L_y,L_z$. 
The initial conditions constructed in this way are slightly different from the
1D profiles returned by MESA models around the iron opacity peak. We
only guarantee that the characteristic physical parameters we choose
are consistent with the MESA stellar models.  In particular, our 
initial condition does not have any
mixing length model of convective or advective energy transport.  We
then calculate how the energy will be transported using
self-consistent 3D radiation hydrodynamic simulations.

At the bottom of the box, we use a reflecting boundary condition for gas
quantities. This means that the density, opacity and the horizontal
components of the velocity are copied from the last active zones to
the ghost zones used for boundary conditions.  The vertical component
of velocity in the ghost zones is set to be the opposite of the
velocity in the last active zones.  The radiation flux is fixed at the
constant value $F_{r,i}$ while the radiation energy density $E_r$ is
integrated to the ghost zones using the diffusion equation.  For the
top boundary, all gas quantities and the radiation flux $\bF_r$ are
copied from the last active zones to the ghost zones. When the
photosphere is included in the simulation box ({\sf StarTop} \&
  {\sf StarMid}), the radiation energy density $E_r$ in the ghost zones is
also copied from the last active zones.  When the whole
simulation box is optically thick ({\sf StarDeep}), $E_r$ is extended
to the ghost zones using the diffusion equation. Periodic boundary
conditions are used for all quantities in the horizontal direction.

For the short characteristic module we use to calculate the VET, the
incoming specific intensities from the bottom of the domain are chosen
such that angular quadratures of the intensities give a constant
vertical flux $F_{r,i}$ and the same $E_r$ as in the moment equations
at the bottom. The incoming specific intensities at the top of the
domain are set to zero when the photosphere is inside the simulation
box.  By contrast, when the whole simulation box is optically thick,
the incoming specific intensities from the top of the domain are set
to be the same as the outgoing specific intensities.

\begin{table}
\hspace{-6cm}
\caption{Simulation Parameters}
\begin{tabular}{cccc}
\hline
Variables/Units					&	{\sf StarDeep}		& {\sf StarMid} 			&	{\sf StarTop} 	\\
\hline
$r_0/\rsun$					&	125.7			&65.6                		&		13.6\\
$T_0$/K						&	$1.87\times 10^5$	&$1.67\times10^5$		&	$1.57\times 10^5$	\\
$\rho_0$/(g cm$^{-3}$)			&	$3.10\times10^{-8}$	&$3.60\times 10^{-9}$	&	$5.52\times 10^{-9}$	\\
$g$/(cm s$^{-2}$)				&	63.59			& $3.56\times 10^2$		&	 $1.17\times 10^{4}$	\\
$F_{r,i}$/(erg cm$^{-2}$ s$^{-1}$)	&	$1.24\times10^{12}$	& $9.94\times 10^{12}$	&	$3.06\times 10^{14}$ 		\\
$H_0$/cm						&	$1.56\times 10^{12}$& $1.52\times 10^{12}$	&	 $2.37\times10^{10}$	\\
$t_0$/s						&	$1.56\times10^{5}$	& $6.54\times 10^{4}$	&	$1.42\times 10^{3}$	\\
$\tau_c\equiv c/c_{g,0}$			&	$5.99\times 10^{3}$	& $6.62\times10^3$		&	$6.54\times10^3$	\\	
$\tau_0$						&	$9.12\times 10^{4}$	& $6.41\times10^3$		& $166.5$		\\	
$P_{r,0}/P_{0}$					&	3.96				& 26.32				&	13.22	\\
$L_{x,y}/H_0$					&	0.89				& 0.46				&	1.17		\\
$L_z/H_0$					&	6.26				& 3.71				&	4.60		\\
$N_{x,y}$						&	128				& 128				&	128	\\
$N_z$						&	768				& 1024				&	512		\\	
\hline
\end{tabular}
\label{Table:parameters}
\begin{tablenotes}
\item    Note: The gravitational acceleration $g$ and radiation flux $F_{r,i} $ coming from the bottom of the simulation box 
are fixed. The optical depth $\tau_0$, temperature $T_0$, density $\rho_0$, radiation and gas pressure $P_{r,0}$, $P_0$ 
are the fiducial parameters for each run. They are also the initial
conditions around the iron opacity peak. The number of cells are
$N_x,N_y$ and $N_z$. 
\end{tablenotes}
\end{table}

\section{Results}
\label{sec:results}
\subsection{The run {\sf StarDeep} when $\tau_0\gg\tau_c$}
The run {\sf StarDeep} is in the regime where the gas and radiation
are tightly coupled and we expect efficient convection.
Below, we first show how the simulated envelope evolves, then
summarize its time averaged structure, and finally compare the energy
transport in our 3D radiation hydrodynamic simulation with mixing
length theory.


\subsubsection{Simulation History}

\begin{figure}[h!]
\begin{center}
\includegraphics[width=1\columnwidth]{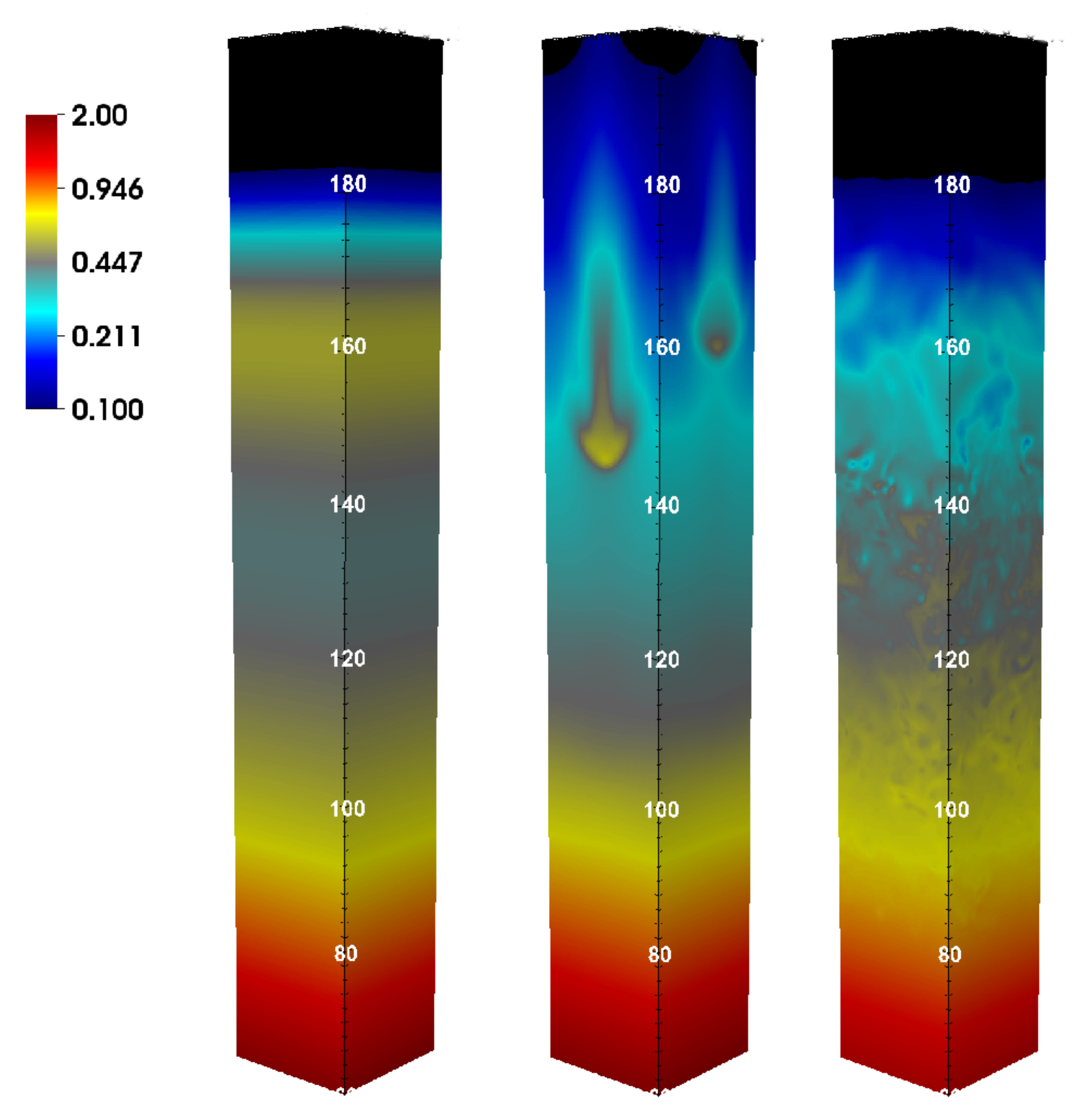}
\caption{Snapshots of density for the run {\sf StarDeep} at time 0 (left), $21.4t_0$ (middle) and 
$97.8t_0$ (right). The box sizes labeled in the plots are in units of solar radius $\rsun$. 
The horizontal box size is $L_x=L_y=20\rsun$. Density is in units
of $\rho_0$, which is given in Table \ref{Table:parameters} for this run.%
}
\label{StarDeepRho}
\end{center}
\end{figure}

Three snapshots of density in Figure \ref{StarDeepRho} show the evolution of 
the envelope. The initial density varies only vertically. At the bottom where 
the temperature is $30\%$ larger than $T_0$, the radiation acceleration is
smaller than the gravitational acceleration and both density and temperature
decrease with height. Around $z=130\rsun$, where we enter the temperature range
of the iron opacity peak, the radiation acceleration becomes larger than the
gravitational acceleration for the fixed radiation flux $F_{r,i}$, and
the density increases with height, peaking around $z=160\rsun$. 
Because the temperature always decreases with height according to the
diffusion equation, the radiation acceleration eventually decreases
after the iron opacity peak.  When the radiation flux again becomes
sub-Eddington, the density drops with height.  The end result is that
the initial condition has the  density inversion associated
with hydrostatic envelopes with a super-Eddington radiation flux
\citep{Jossetal1973}. This profile is our initial condition
for the simulation,
and does not represent the profile in the MESA 1D calculation.

\begin{figure}[h!]
\begin{center}
\includegraphics[width=1.0\columnwidth]{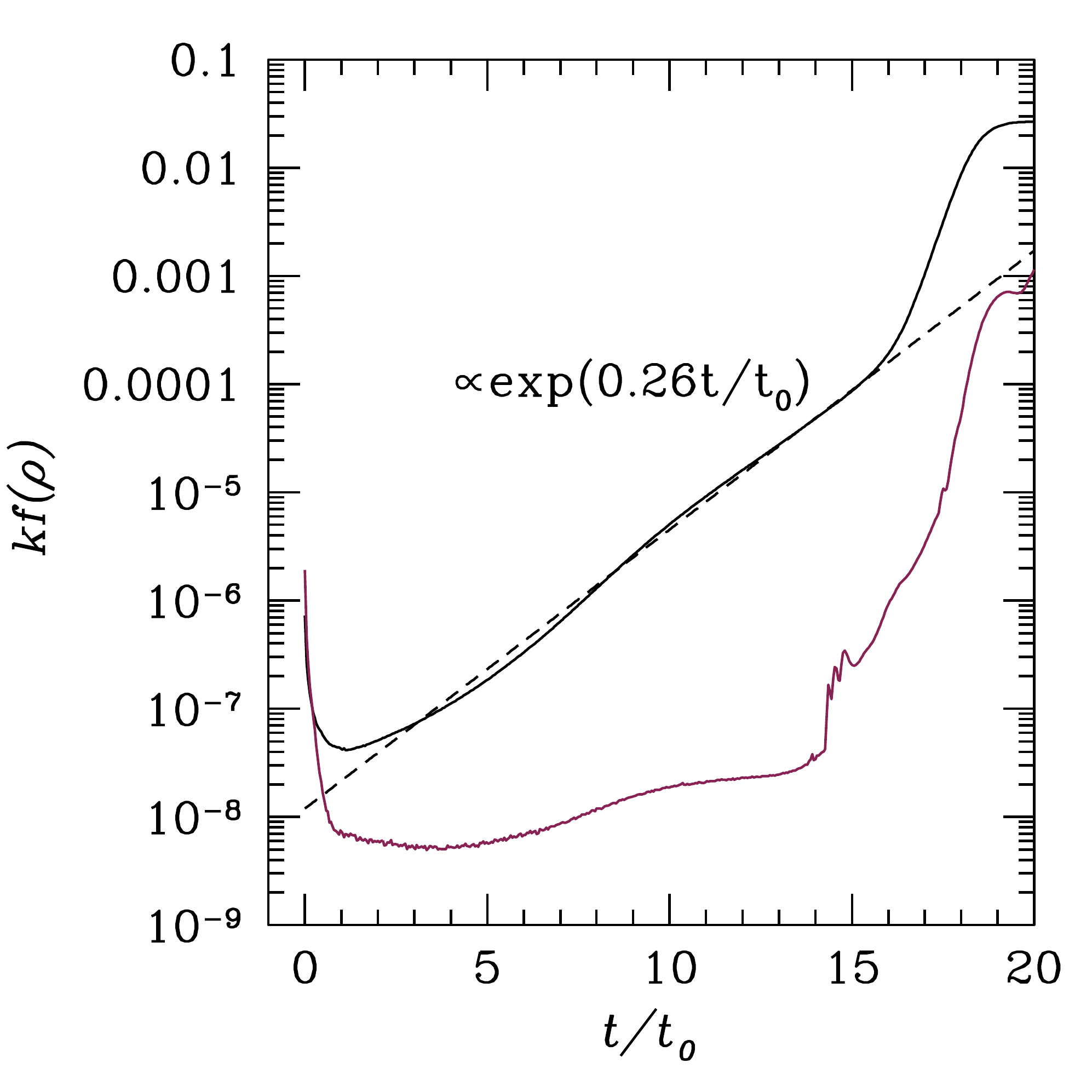}
\caption{Time evolution of the density power spectrum $kf(\rho)$ for the run {\sf StarDeep} 
at height $z=160R_{\odot}$ during the linear growth phase of convection. The black and red
 lines are for the modes with wavelength $0.41H_0$ and $0.18H_0$, respectively. 
The dashed black line is a fit of an exponential function to the solid black line, which 
indicates an e-folding time $3.85t_0$ for this mode. The power spectrum is normalized by 
the mean density at $z=160R_{\odot}$, such that the integral of
$k^2f^2(\rho)$ over all wavenumbers is unity.
}
\label{StarDeepLinear}
\end{center}
\end{figure}

\begin{figure}[h!]
\begin{center}
\includegraphics[width=1.0\columnwidth]{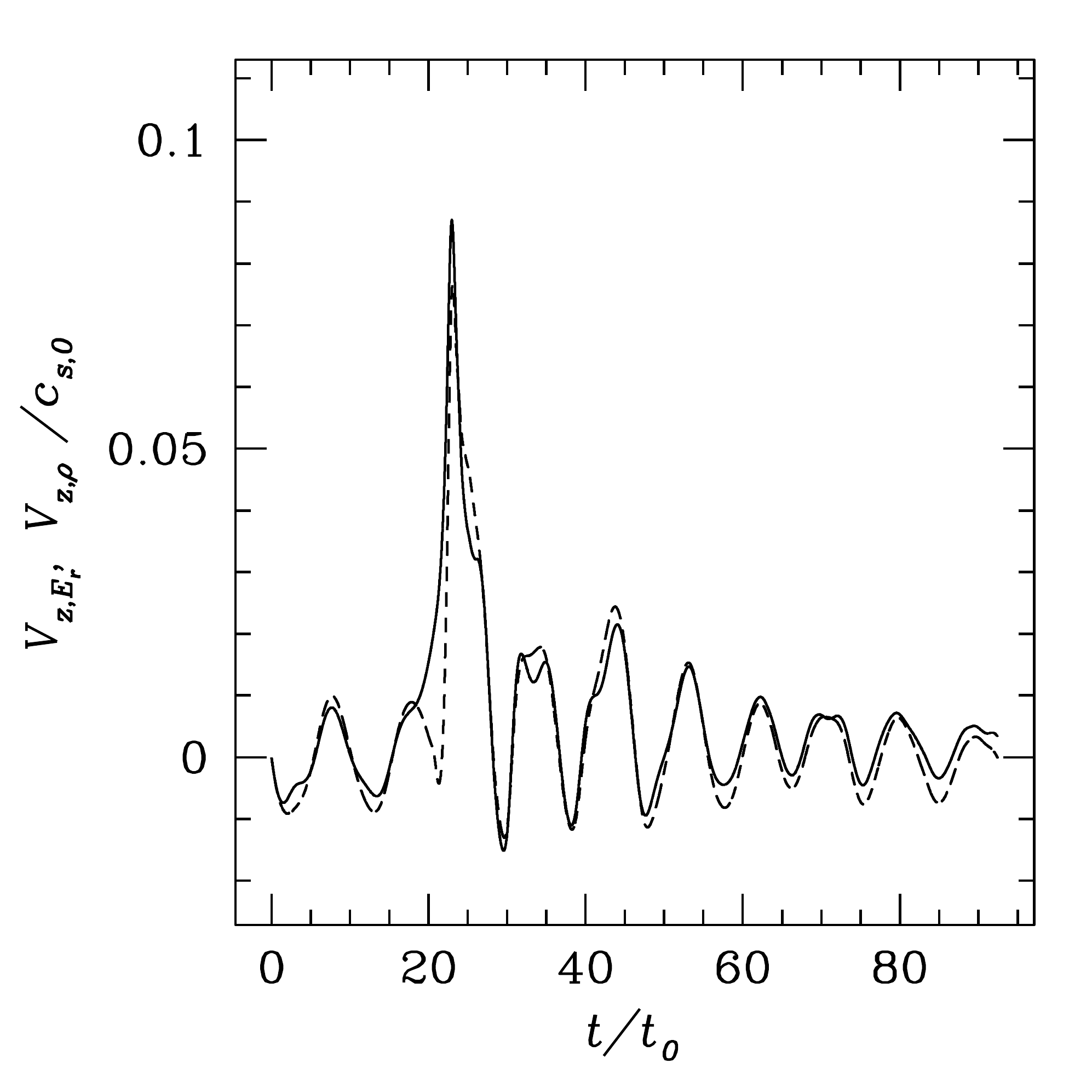}
\caption{History of the radiation energy density weighted ($V_{z,E_r}$, solid line) 
and density weighted ($V_{z,\rho}$, dashed line) vertical velocity for the run {\sf StarDeep}.
}
\label{StarDeepHist}
\end{center}
\end{figure}

The middle panel of Figure \ref{StarDeepRho} shows that the initial
density inversion is unstable.  By $t \approx  20 t_0$, there are large
rising and sinking plumes characteristic of convection and/or the
Rayleigh-Taylor instability.  We now show how this instability can be
quantitatively understood as arising from convection in the radiation
dominated regime.

The gas and radiation entropy per unit mass $S_g,S_r$ are given by 
\begin{eqnarray}
S_g&\equiv&\frac{\kb}{\mu(\gamma-1)}\ln\left[\frac{P/P_{0}}{\left(\rho/\rho_0\right)^{\gamma}}\right]
            \ \ \ {\rm and} \nonumber\\
S_r&\equiv&\frac{4E_r}{3\rho T}.
\label{eq:entropy}
\end{eqnarray}
In our simulations, the total entropy is dominated by the radiation.
This increases with height near the bottom of the simulation
box in the initial condition, but necessarily decreases with height
near the density inversion where $T$ decreases but $\rho$ increases
(see Fig. \ref{StarDeepProfile} discussed below).  We thus expect that
our initial condition is convectively stable near the bottom of the
domain, but convectively unstable near the density inversion region in
3D.  This is exactly what we find in the simulation (see
Fig. \ref{StarDeepRho}).  Convection happens around $160\rsun$ within
$\approx  20t_0$ (equation \ref{eq:tc}) and mixes the initial density
inversion.  High density fingers sink and low density gas rises
upwards.
By $t = 97.8t_0$ (third panel of Figure \ref{StarDeepRho}), the bulk
of the stellar envelope has reached a statistically steady turbulent
state driven by convection.

The initial radiation Brunt-V\"{a}is\"{a}l\"{a} frequency $|N_r|$
(e.g.  equation 52 of \citealt{BlaesSocrates2003}) at the iron opacity
peak is $0.8/t_0$, but it takes $\approx 20t_0$ for convection to begin destroying the initial density
inversion. The wavelength of the fastest growing initial mode is
about half of the horizontal box size, which also differs from the
normal Rayleigh-Taylor instability  with a density discontinuity \citep{Jiangetal2013a} where the
shortest wavelengths grow fastest.  To understand the linear growth
phase of the instability, we first calculate the 2D discrete Fourier
transform in the $x,y$ plane of the density at height $z=160R_{\odot}$
as a function of time. The 2D Fourier coefficient is then binned into
a 1D power spectrum $kf(\rho)$, where $k$ is the magnitude of the
horizontal wavenumber. The time evolution of two modes with horizontal
wavelengths $0.41H_0=9.2R_{\odot}$ and $0.18H_0=4.0R_{\odot}$ are
shown in Figure \ref{StarDeepLinear}. The decrease of $kf(\rho)$
within $t_0$ probably originates from an initial vertical
oscillation that is present due to the imperfect balance between
gravity and pressure gradients in the initial condition. After this
point, the long wavelength mode grows exponentially with an e-folding
time $3.85t_0$ until $15t_0$.  The shorter wavelength mode also
grows exponentially, but at a rate that is smaller by a factor of
$3.44$ initially.  However, it then grows quickly after
$15t_0$ as convection becomes nonlinear and power from long
wavelength modes cascades to the short wavelength modes.

The measured growth rates agree very  well with the short wavelength
WKB linear analysis done by \cite{BlaesSocrates2003} (the last factor
of their equation 59). Plugging in numbers for the long wavelength
mode in Figure \ref{StarDeepLinear}, and assuming that the vertical wavelength of this mode is about
equal to its horizontal wavelength, we get a growth rate of
$0.27/t_0$, very close to the value measured in the simulation. 
This decline should only be valid for wavelengths with
growth rates less than the magnitude of the Brunt-V\"{a}is\"{a}l\"{a}
frequency $|N_r|$, giving an expected transition wavelength  
as defined in \cite{BlaesSocrates2003} of roughly
$0.5H_0$ for this simulation.  Hence we expect the longest wavelength
convective modes to be only weakly affected by radiative damping,
which explains why the measured long wavelength growth rate of
$0.26/t_0$ is of the same order as $|N_r|$.  Shorter wavelength modes
will be more strongly affected, and grow much more slowly.  The
measured growth rate of the shorter wavelength mode here is consistent
with a reduction factor $\propto k^{-1.5}$.  This is consistent with
the expected behavior $k^{-2}$, given that we are not too far from the
transition wavelength, and that the short wavelength mode could have a
more horizontal wave vector orientation.

The envelope of the {\sf StarDeep} simulation shows significant vertical
oscillatory motion.  We calculate two measures of the
time-dependent vertical velocity, weighted by energy and by density,
\begin{eqnarray}
V_{z,E_r}=\frac{\int v_zE_r d V}{\int E_r d V}, V_{z,\rho}=\frac{\int v_z\rho d V}{\int \rho d V}, 
\end{eqnarray}
where the integrals are done over the whole simulation box.
Histories of these two vertical velocity measures are shown in
Figure \ref{StarDeepHist}. The averaged density, or equivalently the 
total mass, drops by $9\%$ around $20t_0$. At the same time,
$V_{z,E_r}$ and $V_{z,\rho}$ 
reach their peak values. This corresponds to the time when 
the initial density inversion is destroyed due to convection. 
After the initial transient, the envelope slowly settles down to a steady state, and 
both $V_{z,E_r}$  and $V_{z,\rho}$ show oscillations with amplitudes
of $1-2\%$ of the radiation sound speed. 
The average density also slowly decreases with time because of the mass loss through our open top boundary. 
We have run the simulation for more than five thermal times to reach the thermal equilibrium.

\subsubsection{Time Averaged Vertical Structures}

\begin{figure}[h!]
\begin{center}
\includegraphics[width=1.0\columnwidth]{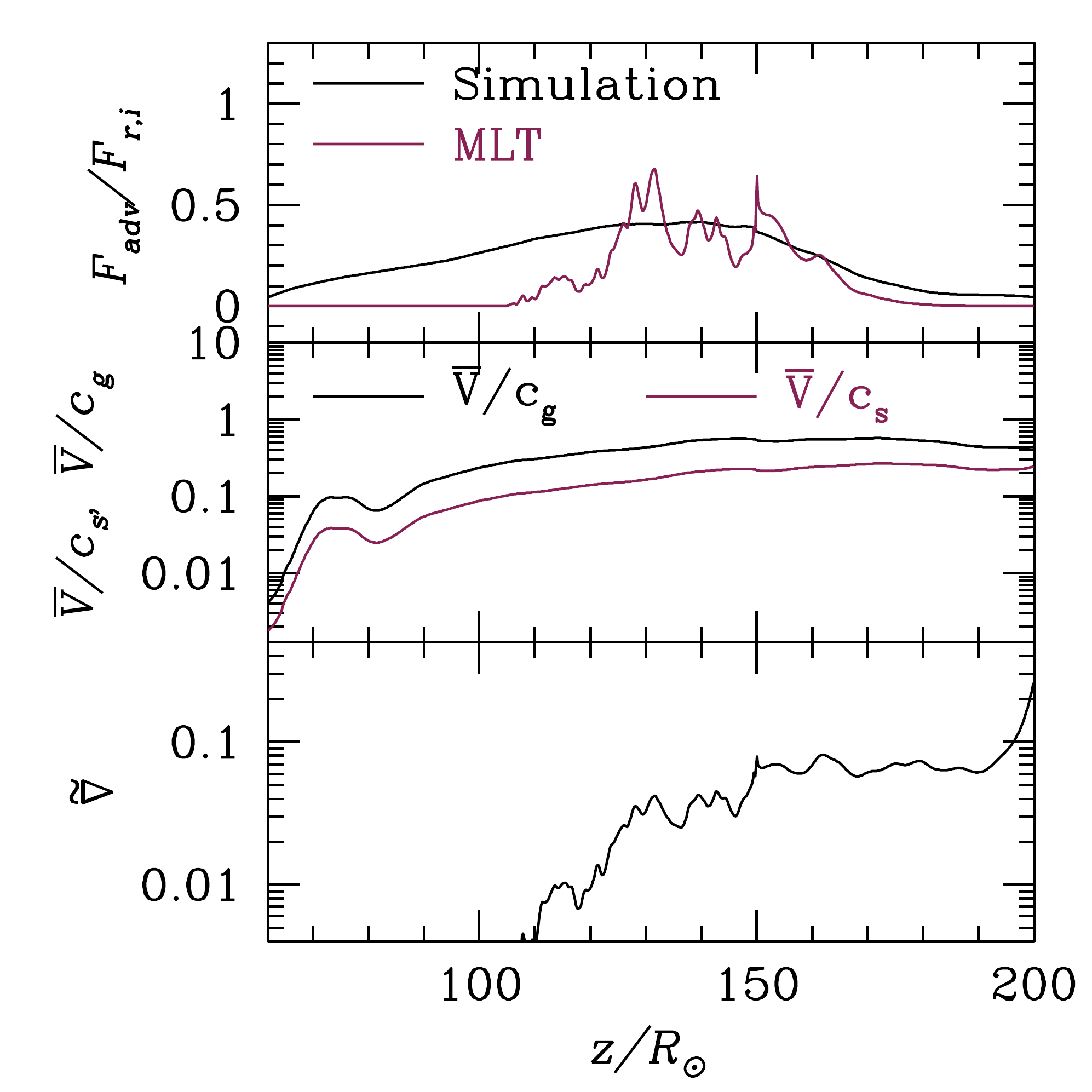}
\caption{Top: horizontally and time averaged advection flux $F_{\rm adv}=v_zE_r$ (black line) for the run 
{\sf StarDeep}. The red line is the convection flux calculated according to equation (\ref{eq:mlt}) 
based on the horizontally and time averaged vertical profiles of temperature and pressure in the same 
run with mixing length parameter $\alpha=0.55$. Middle: horizontally and time averaged turbulent velocity 
$\bar{V}$ (equation \ref{eqn:turbv}), scaled with radiation sound speed $c_s$ (red line) and isothermal 
sound speed $c_g$ (black line) at each height. Bottom: vertical profile of $\tilde{\nabla}\equiv(\nabla-\nabla_{\text{ad}})/\nabla_{\text{ad}}$ 
for the run {\sf StarDeep}. The actual gradient $\nabla\equiv d\ln T/d\ln(P+P_r)$ 
is calculated based on the horizontally and time averaged temperature and pressure profiles. The adiabatic 
gradient is close to $0.25$ for the radiation dominated envelope.  The fact that $\tilde{\nabla}\ll 1$ 
implies that the convection is almost adiabatic.  }
\label{StarDeepConvection}
\end{center}
\end{figure}

\begin{figure*}[htp]
\begin{center}
\includegraphics[width=1.0\columnwidth]{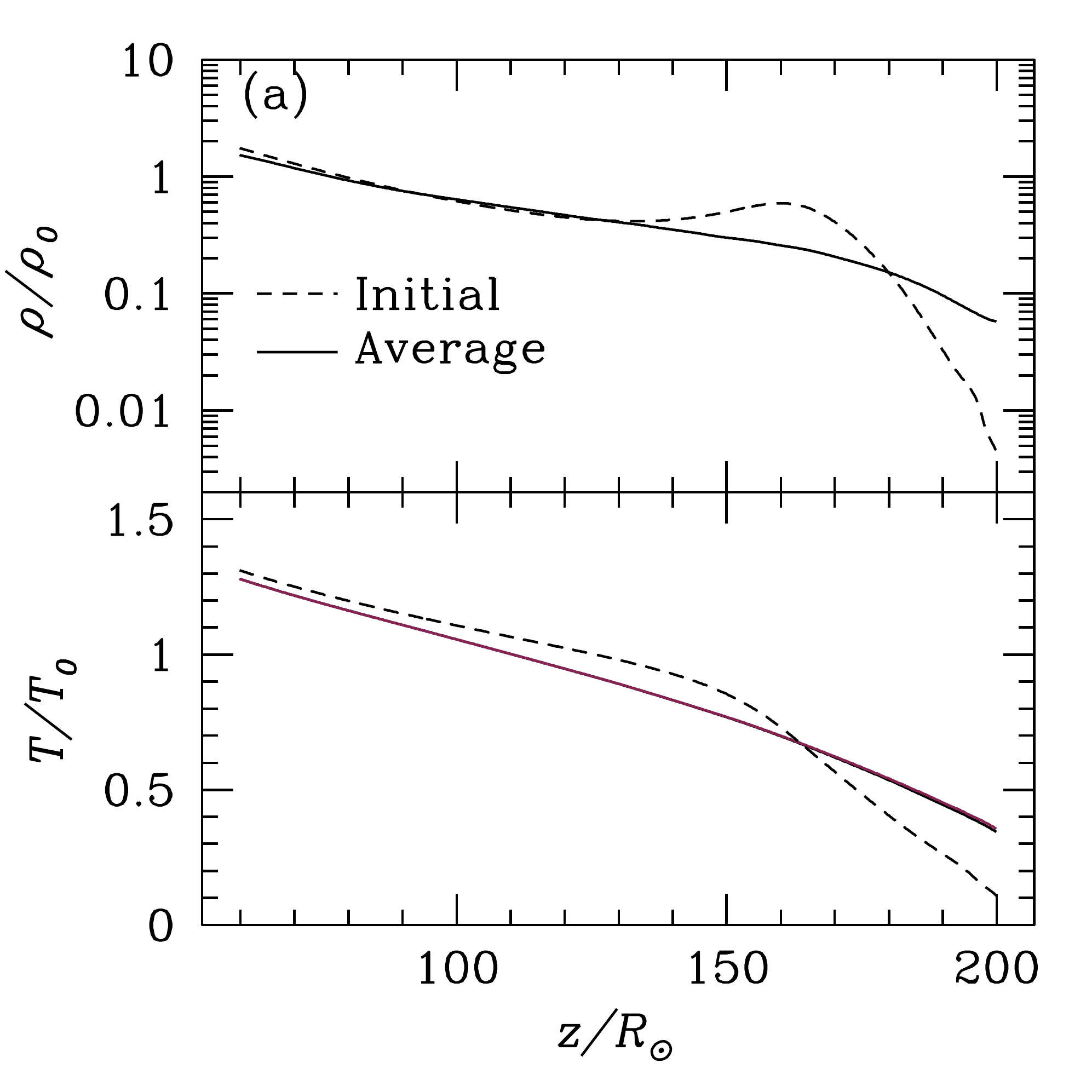}
\includegraphics[width=1.0\columnwidth]{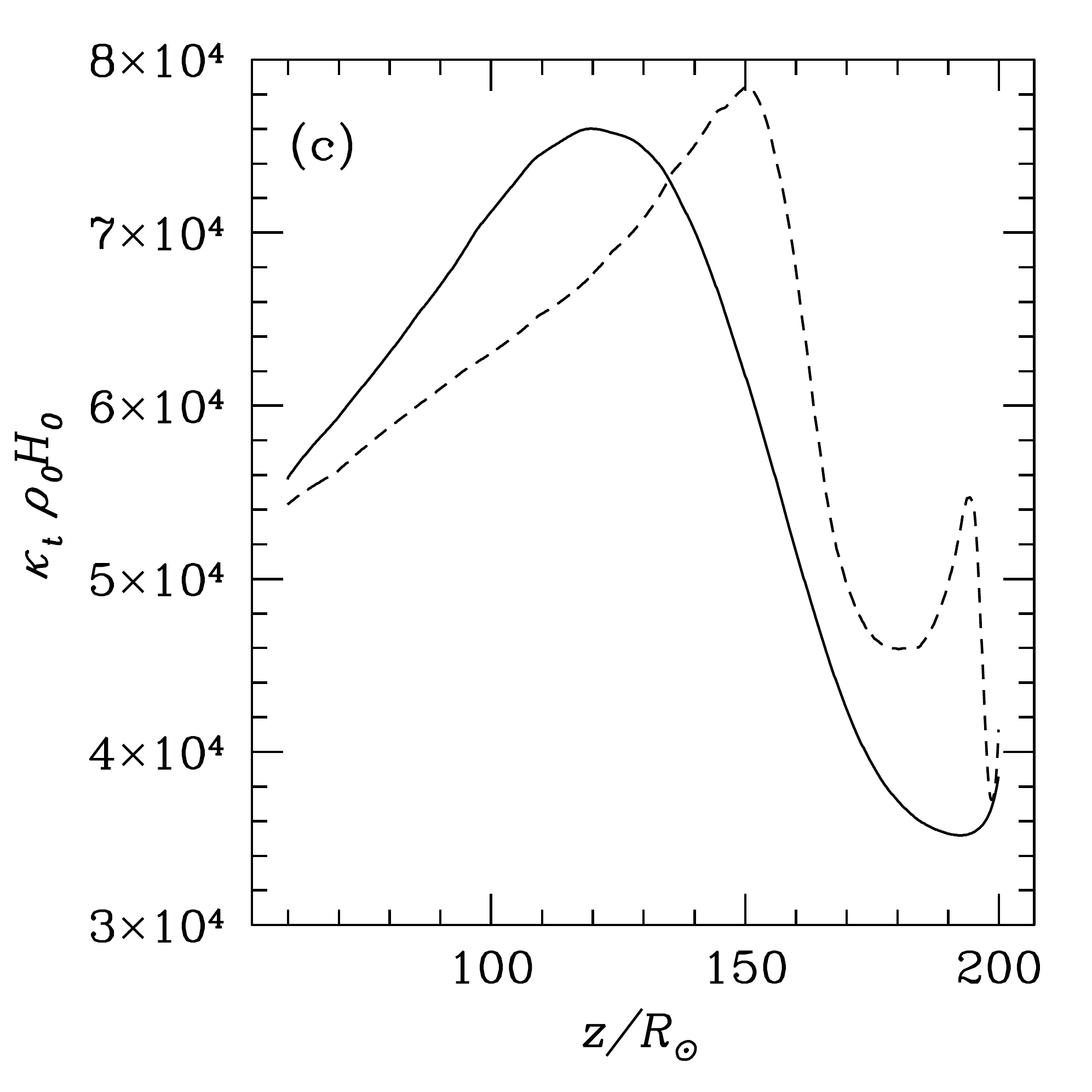}\\
\includegraphics[width=1.0\columnwidth]{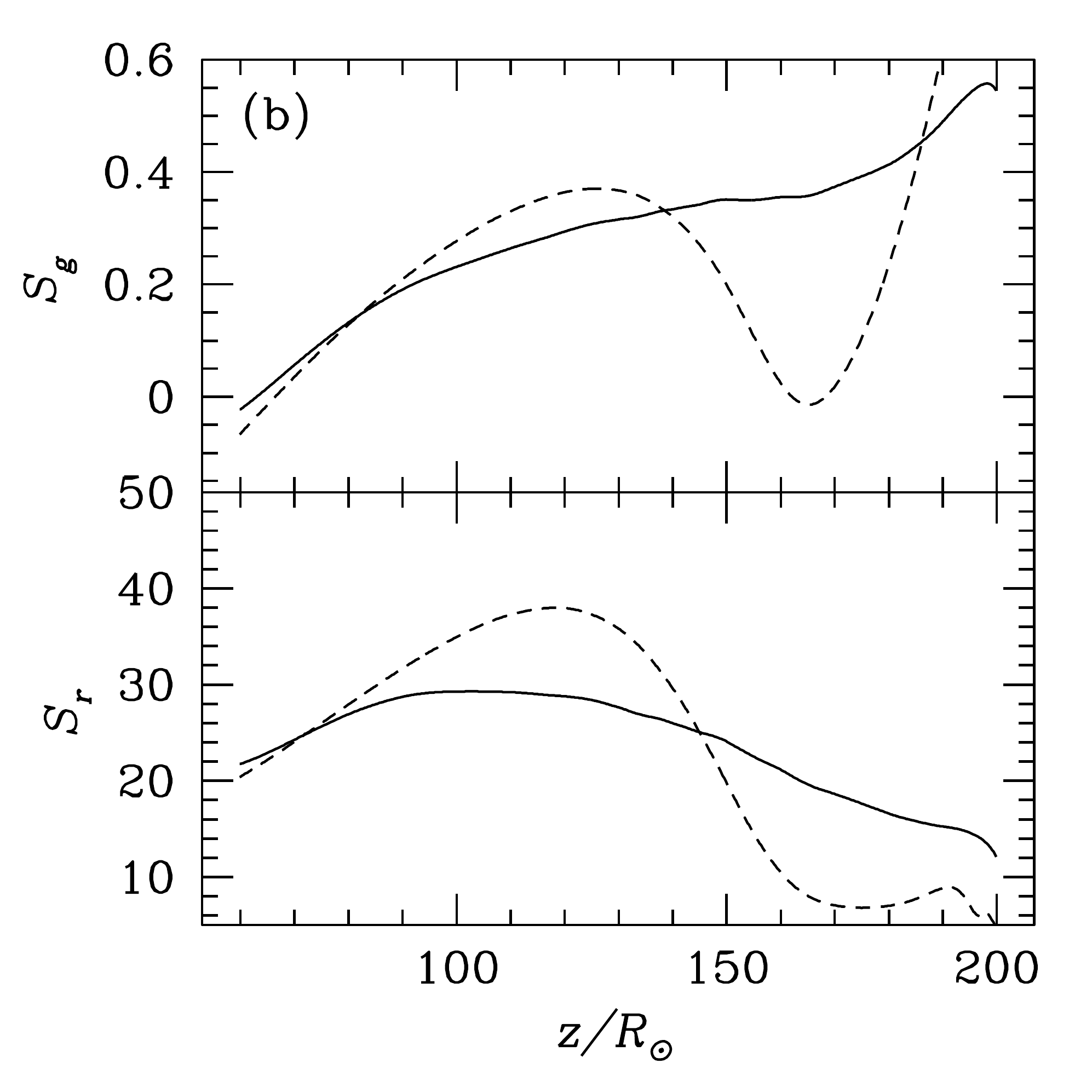}
\includegraphics[width=1.0\columnwidth]{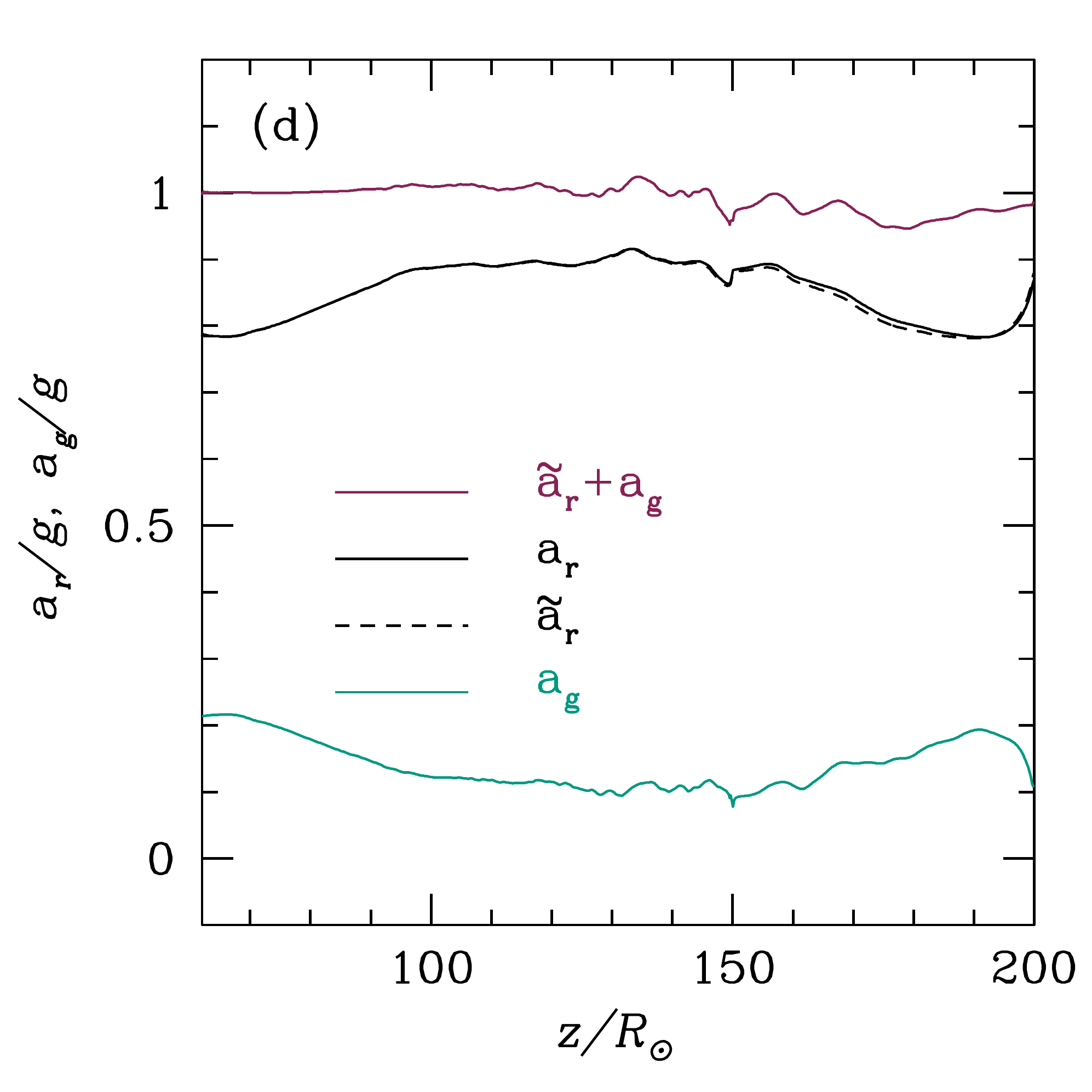}
\caption{Horizontally and time averaged vertical profiles of various quantities for  
the run {\sf StarDeep}. Panel (a): density (solid line at top), gas temperature (solid black 
line at bottom) and radiation temperature (solid red line at bottom). The gas and radiation 
temperature profiles overlap as they are strongly thermally coupled. 
Panel (b): gas entropy $S_g$ (solid black line at top) and radiation 
entropy $S_r$ (solid black line at bottom). 
These entropies are calculated according 
to equation (\ref{eq:entropy}), and both are in units of $\kb/\mu$.
Panel (c): total opacity $\kappa_t$ in unit of $1/(\rho_0H_0)$ (the solid line). 
The dashed black lines in these three panels are the initial 
conditions. Panel (d): volume averaged (solid black line)
and density weighted (dashed black line) radiation accelerations ($a_r$, $\tilde{a}_r$) as 
well as the acceleration due to the gas pressure gradient ($a_g$, green line). 
The red line is the sum of $\tilde{a}_r$ and $a_g$. In this run with 
$\tau_0\gg\tau_c$, the density inversion in the initial condition is
gone. 
}
\label{StarDeepProfile}
\end{center}
\end{figure*}

In the 3D radiation hydrodynamic simulations, energy can be
transported vertically by either photon diffusion or turbulent
transport.  We quantify the turbulent transport using the advection
flux $F_{\rm adv}$, which is
\begin{eqnarray}
F_{\rm adv}=v_z E_r. 
\end{eqnarray}
In principle, photon advection can be due to either a mean flow (e.g.,
an outflow) or turbulence.  In our simulations, the latter dominates.

The solid line in the top panel of Figure \ref{StarDeepConvection}
shows the time averaged advective radiation flux $F_{\rm adv}$ as a
function of height where the time average is done between $31t_0$ and
$107t_0$.  Most of the flux at $z<100\rsun$ is carried by radiative
diffusion.  In the turbulent region between $z=110\rsun$ and
$160\rsun$, however, the averaged advection flux is more than $40\%$
of $F_{r,i}$.  We compare this to MLT in the next section.

The turbulent velocity in the envelope can be quantified as
\begin{eqnarray}
\overline{V}=\sqrt{\frac{\int \rho \bv\cdot\bv d V}{\int \rho d V}}.
\label{eqn:turbv}
\end{eqnarray}
The middle panel of Figure \ref{StarDeepConvection} shows the time
averaged vertical profile of this turbulent velocity relative to both
the isothermal sound speed $c_g=\sqrt{P/\rho}$ and the radiation sound
speed $c_s=\sqrt{E_r/(3\rho)}$ .  The averaged turbulent velocity is
only $ 6-20\%$ of the radiation sound speed in the turbulent
region. Even compared with the isothermal sound speed alone,
$\overline{V}/c_g$ is smaller than $0.4$.  This implies that in the
efficient convection regime, the turbulent kinetic energy density is
much smaller than the thermal energy density ($\lesssim2\%$), as
expected.

Figure \ref{StarDeepProfile} shows the quasi-steady envelope structure
compared with the hydrostatic initial condition.  Note that the
density and temperature are relatively unaffected near the bottom of
the box indicating that we have placed the bottom boundary deep
enough. In the final quasi-steady envelope,
the iron opacity peak moves slightly deeper relative to the initial
condition, but the density inversion present in the initial condition
is gone.  This is consistent with the 1D MESA models which also find
that for the conditions in {\sf StarDeep} mixing length theory
predicts that convection is efficient and is able to remove the
density inversion.
Note that the location where the advection flux is largest
($z\approx 120 R_\odot$ in Fig. \ref{StarDeepConvection}) corresponds to
the location of the iron opacity peak in the final state.  This is
also where the radiation entropy $S_r$ decreases with height,
consistent with convection continuing to drive the turbulent motions
and energy transport.  As expected, the radiation entropy profile is
also flatter in the
turbulent state after the density inversion is removed.  

Because part of the energy flux in the turbulent state is carried by
advection, the averaged diffusive flux also drops below the initial
value $F_{r,i}$. 
The significant contribution of the advection flux reduces the
radiation acceleration so that it is smaller than the gravitational
acceleration.  The accelerations due to gas $a_g$ and radiation $a_r$
can be calculated as
\begin{eqnarray}
a_g=-\frac{1}{\rho}\frac{\partial P}{\partial z},\nonumber\\
a_r=\frac{\kappa_t}{c} F_{r,0z}.
\label{eq:acce}
\end{eqnarray}
Notice that only the diffusive radiation flux $F_{r,0z}$ contributes to the radiation acceleration, which 
is the radiation pressure gradient in the optically thick regime. 
The last panel of Figure \ref{StarDeepProfile} shows that gravity is roughly balanced by 
the sum of radiation and gas pressure gradients, with $a_r$ being at most
$90\%$ of $g$. The fact that $a_r \lesssim g$ is consistent with the
absence of a density inversion in the turbulent state.

\begin{figure*}[htp]
\begin{center}
\includegraphics[width=2.0\columnwidth]{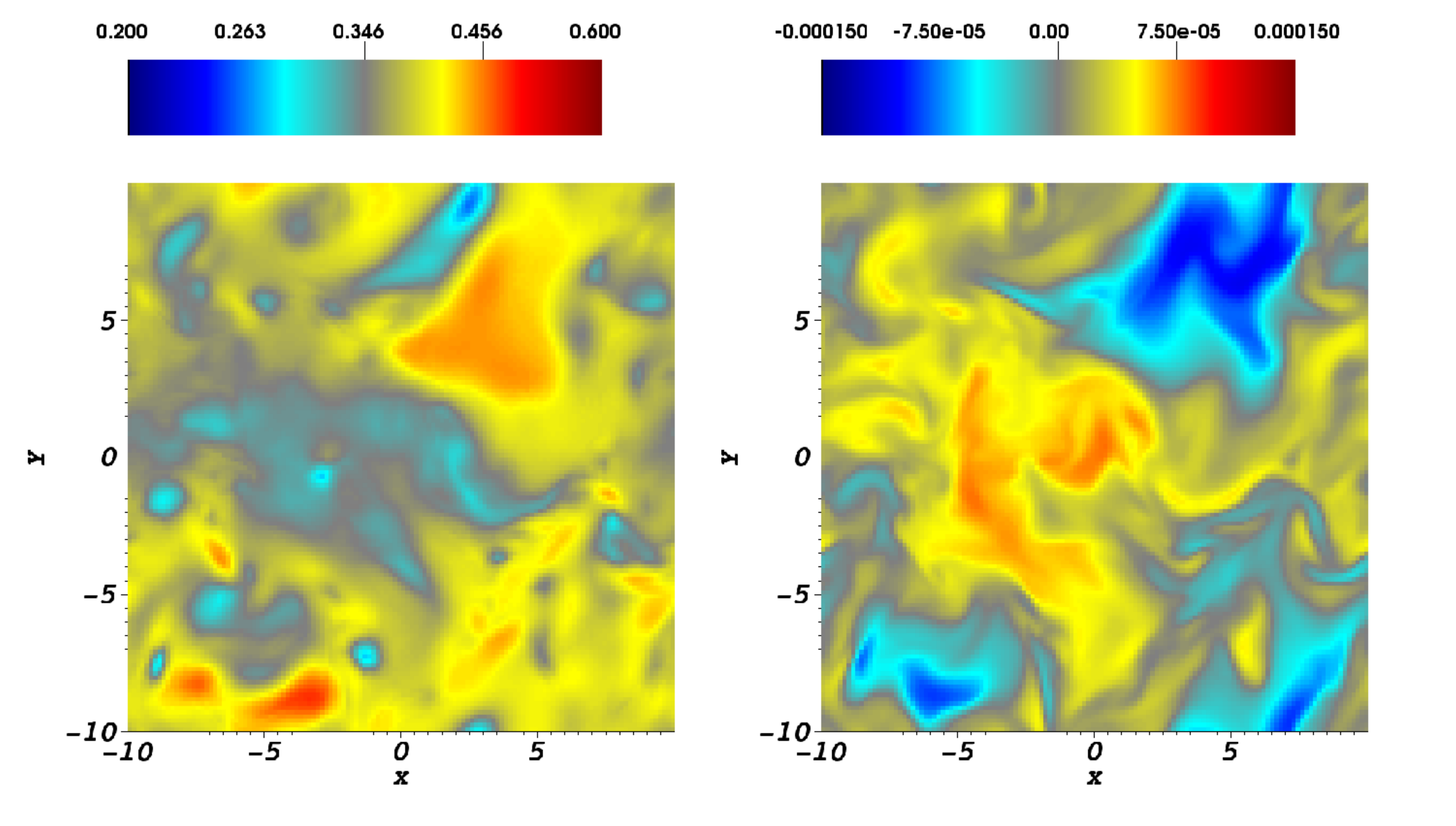}
\caption{Horizontal slices of density $\rho$ (left) and vertical component of the radiation 
flux $F_{r,z}$ (right) at $z=140\rsun$ for the run {\sf StarDeep} at time $97.8t_0$. 
The box size is in units of $\rsun$ while units for $\rho$ and $F_{r,z}$ are 
$\rho_0$ and $ca_rT_0^4$ respectively. Notice the strong anti-correlation between 
$\rho$ and $F_{r,z}$ caused by buoyancy.%
}
\label{StarDeepslice1}
\end{center}
\end{figure*}

\begin{figure}[h!]
\begin{center}
\includegraphics[width=1.0\columnwidth]{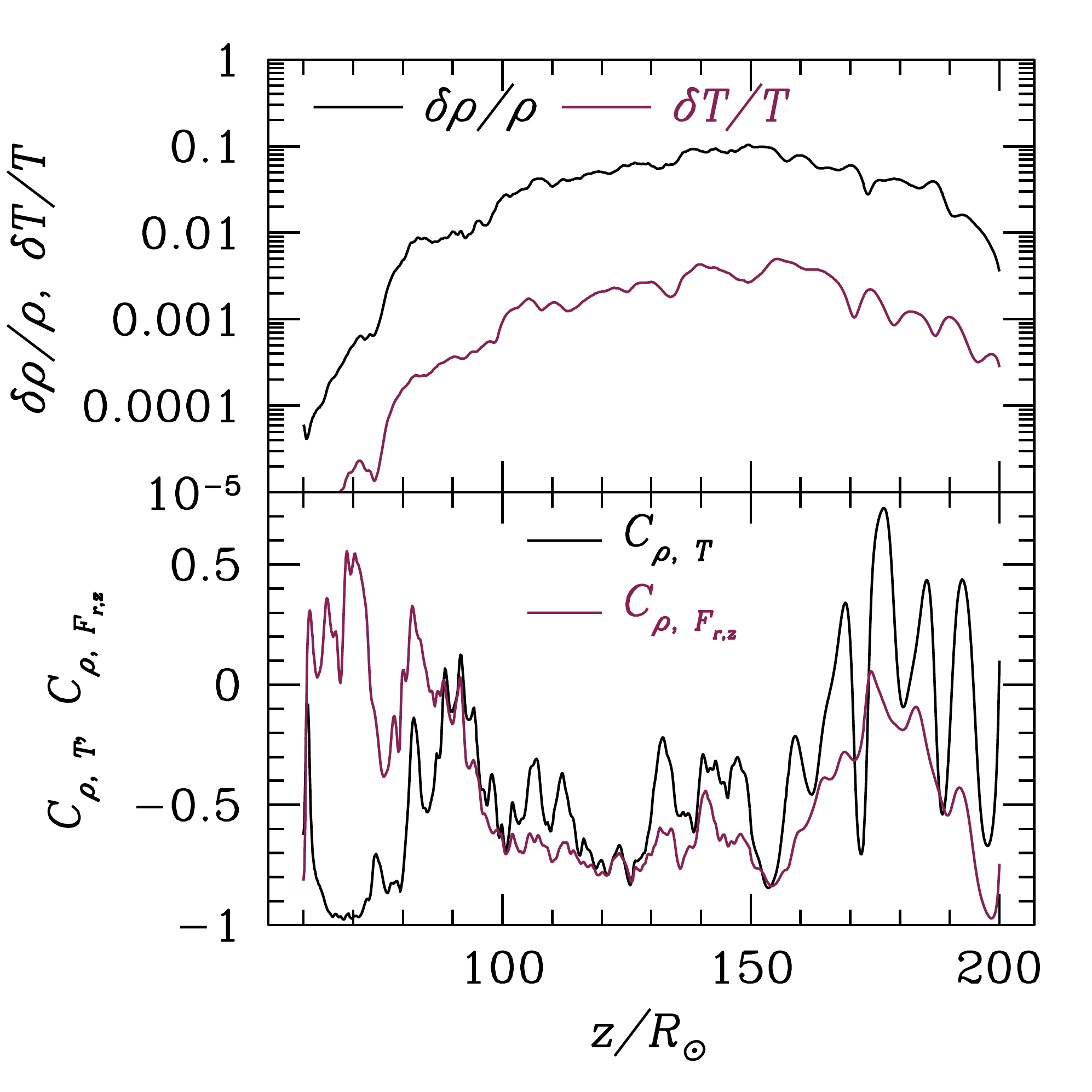}
\caption{Top: vertical profiles of standard deviations of density $\rho$ 
and temperature $T$ at time $97.8t_0$ for the run {\sf StarDeep}. 
The standard deviations are scaled with the mean density and temperature 
at each height. Bottom: Vertical profiles of cross correlation coefficients between 
$\rho$, $T$ and $\rho$, $F_{r,z}$ at the same time.  The anti-correlation 
between $\rho$ and $F_{r,z}$ in this run is due to the buoyancy instead of 
the ``porosity" effect, as $F_{r,z}$ 
is dominated by the advection flux $v_zE_r$. %
}
\label{StarDeepsigma}
\end{center}
\end{figure}

\subsubsection{Time Averaged Horizontal Structures}
\label{sec:HorizStarDeep}


Figure \ref{StarDeepslice1} shows horizontal slices of density $\rho$
and the vertical component of Eulerian radiation flux $F_{r,z}$ at
$z=140\rsun$ and time $97.8t_0$ for the run {\sf StarDeep}.  There is
a clear anti-correlation between fluctuations of $\rho$ and $F_{r,z}$.
When $\rho$ is larger (smaller) than the horizontally averaged
density, $F_{r,z}$ is negative (positive). 
This is a natural outcome of convection, where the 
Eulerian radiation flux $F_{r,z}$ is dominated by the
advection part $v_zE_r$ at each location in the regime
$\tau_0\gg \tau_c$. Figure \ref{StarDeepslice1} thus shows that low
density regions buoyantly rise while high density fluid elements fall.

To quantify the level of density and temperature fluctuations, the top
panel of Figure \ref{StarDeepsigma} shows the vertical profiles of
standard deviations of $\rho$ and $T$ scaled with the horizontally
averaged mean values at each height at time $97.8t_0$.  The scaled
standard deviations peak at $150\rsun$, where the iron opacity
peak is located at this time. Beyond the convective region,
fluctuations drop with height. The scaled density standard deviation
is always smaller than $10\%$ in this case. Temperature fluctuations
($\lesssim 0.4\%$) are much smaller.  Because radiation pressure
$P_r\propto T^4$, the pressure fluctuations are only $\lesssim 1.6\%$.

The cross correlation between two quantities $a$ and $b$ can be
quantified by the correlation coefficient $C_{a,b}$, normalized by
their standard deviations $\sigma_a,\sigma_b$.  The bottom panel of
Figure \ref{StarDeepsigma} shows the correlation coefficients between
$\rho,T$ and $\rho,F_{r,z}$. Consistent with the slice shown in Figure
\ref{StarDeepslice1}, there is an anti-correlation between $\rho$ and
$F_{r,z}$ in the convective region due to buoyancy. Temperature
fluctuations also anti-correlate with density fluctuations, which
means an anti-correlation between density $\rho$ and radiation
pressure $P_r$. In this very optically thick regime $\tau_0\gg\tau_c$,
density fluctuations do not reduce the effective radiation
acceleration, which can be confirmed by comparing the volume averaged
$a_r$ (equation \ref{eq:acce}), and the density weighted radiation
acceleration
\begin{eqnarray}
\tilde{a}_r=\frac{\langle\rho\kappa_tF_{r,0z}\rangle}{c\langle\rho\rangle}.
\label{eq:artilde}
 \end{eqnarray}
 Here the average $\langle\cdot\rangle$ is done horizontally at each
 height.  Panel (d) of Figure \ref{StarDeepProfile} shows vertical
 profiles of the time averaged radiation accelerations calculated in
 these two different ways.  They are almost identical.  This explictly
 demonstrates that there is no reduction of the radiation acceleration
 due to density fluctuations in this regime.

 The typical size of the turbulent structures can be estimated by
 calculating the normalized cross correlation $C_{\rho,\tilde{\rho}}$
 between $\rho(x,y)$ and $\rho(x+l_{\rho},y+l_{\rho})$, where
 $\rho(x+l_{\rho},y+l_{\rho})$ is the density shifted horizontally
 with distance $l_{\rho}$.  The correlation length $l_{\rho}$ can be
 defined to to be the shifted distance when $C_{\rho,\tilde{\rho}}$
 drops to zero. For the run {\sf StarDeep}, $l_{\rho}$ is typically
 $\sim 0.3-0.4H_0$, with corresponding optical depth at the iron
 opacity peak comparable to $c/c_{s,0}$ given in Table
 \ref{Table:parameters}.

\subsubsection{Comparison with Mixing Length Theory}

In the radiation dominated regime, the radiation advection flux
$F_{\rm adv}$ from the simulations can be directly compared with the
convection flux from MLT.  We define the ratio of gas pressure
to total pressure ($\beta$), the first ($\Gamma_1$) and second
($\Gamma_2$) adiabatic indices, and specific heat at constant
pressure ($c_p$) for a mixture of gas and radiation
\citep[][]{Chandra1967}, and the logarithmic derivative of density
with respect to temperature at constant total pressure to be
\begin{eqnarray}
\beta&\equiv&\frac{P}{P+P_r},\nonumber\\
\Gamma_1&\equiv&\beta+\frac{(4-3\beta)^2(\gamma-1)}{\beta+12(\gamma-1)(1-\beta)},\nonumber\\
\Gamma_2&=&\frac{(4-3\beta)\Gamma_1}{3(1-\beta)\Gamma_1+\beta}\nonumber\\
c_p&=&\frac{1}{\gamma-1}\frac{\Gamma_1}{\beta^2}\left[
\beta+12(\gamma-1)(1-\beta)\right],\nonumber\\
Q&\equiv&-\left.\frac{d\ln \rho}{d\ln T}\right|_{P+P_r}=\frac{4-3\beta}{\beta},
\end{eqnarray}
where $P_r$ is the $zz$ component of the radiation pressure tensor ${\sf P_r}$. 
The adiabatic temperature gradient with respect to total pressure can be calculated as
\begin{eqnarray}
\nabla_{\text{ad}}\equiv\left.\frac{d\ln T}{d\ln(P+P_r)}\right|_{\text{ad}}&=&\frac{\Gamma_2-1}{\Gamma_2}.
\end{eqnarray}
The actual gradient $\nabla\equiv d\ln T/d(\ln(P+P_r))$ can be
calculated based on the mean density and temperature profiles shown in
Figure \ref{StarDeepProfile}. If we assume that the mixing length is
related to the pressure scale height via the mixing parameter
$\alpha$, then the convective flux based on non-adiabatic mixing length theory including 
the effects of radiative cooling is \citep[][]{Henyey:1965}
\begin{eqnarray}
 F_{\text{conv}}=\frac{Q^{1/2}c_p}{4\sqrt{2}}\frac{\kb\rho T}{\mu}\left(\frac{P+P_r}{\rho}\right)^{1/2}\alpha^2\left(
 \nabla-\nabla'\right)^{3/2},\nonumber\\
 \label{eq:mlt}
 \end{eqnarray}
 assuming that the radiation entropy $S_r$ decreases with height.  
 Here the gradient $\nabla'$ is calculated based on 
equations (32) and (42)\footnote{Notice that there is a typo in equation (42) of 
\cite{Henyey:1965}. The $\alpha$ should be $\alpha^2$. We thank Jeremy Goodman 
for pointing this out.} in \cite{Henyey:1965} given the mean 
vertical profiles of the envelope. In the adiabatic limit, $\nabla'$ is reduced 
to $\nabla_{\text{ad}}$. Then equation \ref{eq:mlt} is the same as the formula 
given by the adiabatic mixing length theory \citep[][]{Cox1968}.

 The convective flux calculated in this way with $\alpha=0.55$ is
 shown as the red line in the top panel of Figure
 \ref{StarDeepConvection}, which agrees with the time averaged
 advection flux in the simulation. 
 Note that the value of $\alpha$
 required to explain the flux in the simulations is also quite close
 to the ratio between the density correlation length and pressure
 scale height $H_0$ (see Section \ref{sec:HorizStarDeep}), which is
 the definition of $\alpha$ in mixing length theory.  Finally, Figure
 \ref{StarDeepConvection} shows that the difference between the actual
 gradient and the adiabatic gradient
 $\tilde{\nabla}\equiv (\nabla-\nabla_{\text{ad}})/\nabla_{\text{ad}}$
 is quite small, consistent with the expectations of efficient
 convection.

 Taken together, these comparisons demonstrate that the statistical
 properties of the turbulent energy transport in our radiation
 hydrodynamic simulation with $\tau_0 \gg \tau_c$ are reasonably
 consistent with the expectations of mixing length theory, although
 the value of $\alpha$ is somewhat smaller than traditionally assumed
 in models of massive stars \citep[typical values $\alpha= 1.0...1.5$,
   see e.g.][]{Yusof2013,Koheler2015}. This is the first numerical
 confirmation of the adequacy of MLT in a radiation dominated regime. 


\subsection{The run {\sf StarMid} when $\tau_0\gtrsim\tau_c$}
In this regime, rapid radiative diffusion starts to have a
non-negligible impact on the properties of convection in the radiation
dominated stellar envelope (the typical thermal time scale near the
iron opacity peak is $\sim 5$ dynamical times).  Compared with the
previous run, we shall see the effects of the reduced optical depth.
Note that unlike {\sf StarDeep}, the photosphere is captured within the
domain of this simulation. 

\subsubsection{Simulation History}

\begin{figure}[h!]
\begin{center}
\includegraphics[width=1.0\columnwidth]{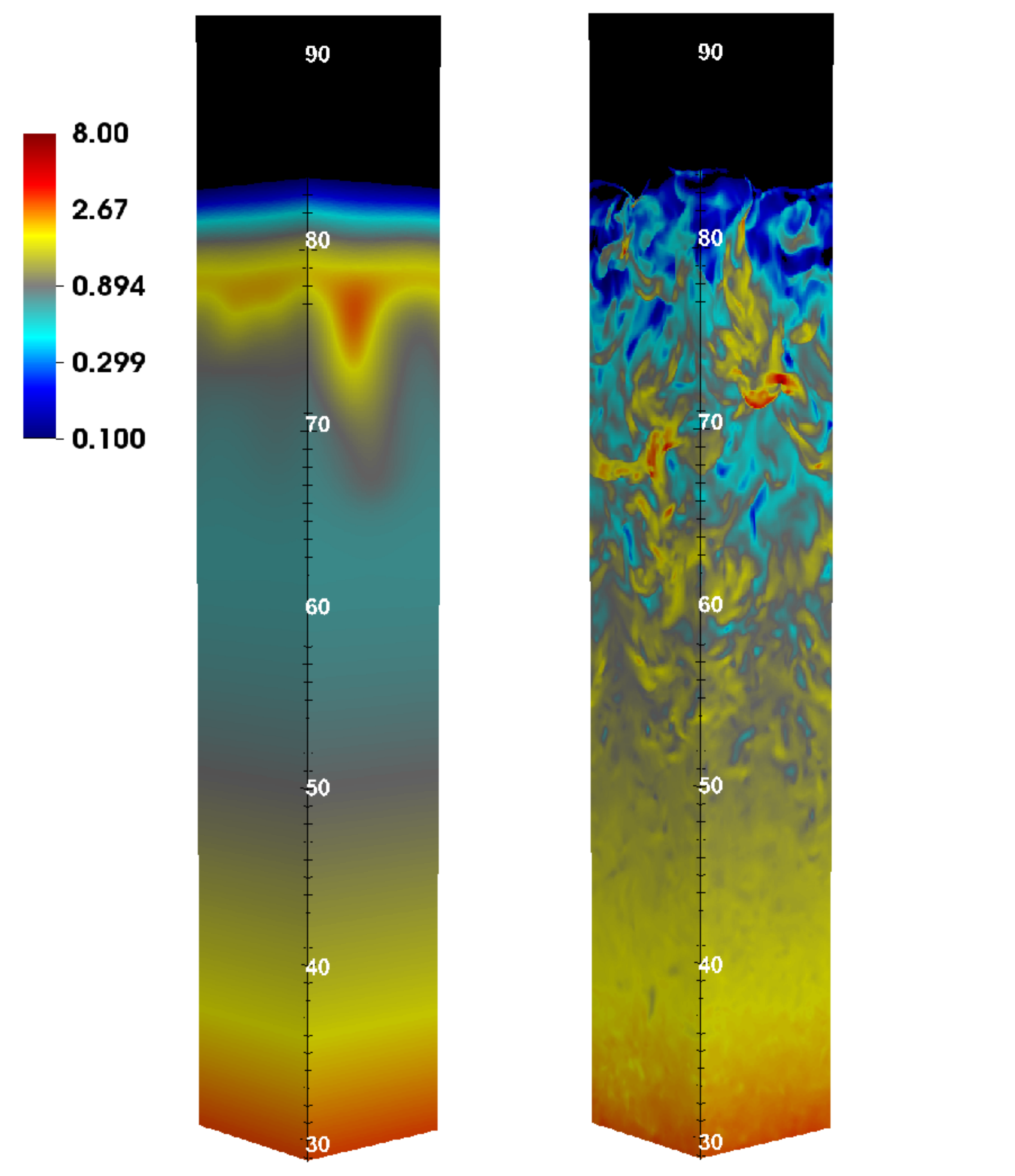}
\caption{Snapshots of density for the run {\sf StarMid} at time $9.4t_0$ (left) and 
$18.8t_0$ (right). The box size is in units of $\rsun$ while the density unit $\rho_0$ 
is given in Table \ref{Table:parameters}. The horizontal box sizes are $L_x=L_y=10\rsun$.%
}
\label{StarMidRho}
\end{center}
\end{figure}

\begin{figure}[h!]
\begin{center}
\includegraphics[width=1.0\columnwidth]{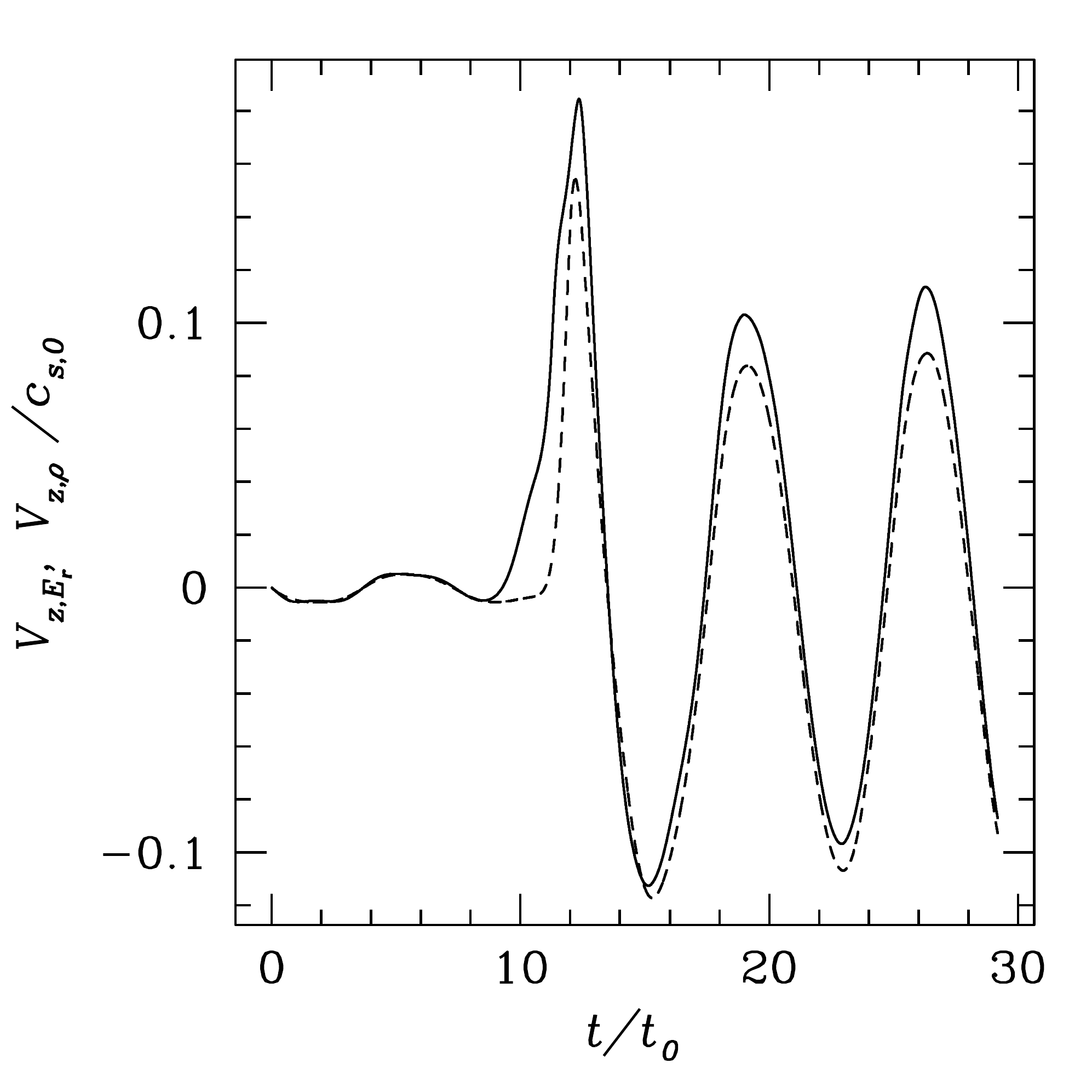}
\caption{History of the volume averaged vertical velocities $V_{z,E_r}$ and $V_{z,\rho}$ for the run {\sf StarMid}. 
}
\label{StarMidHist}
\end{center}
\end{figure}

The initial condition for {\sf StarMid} has the iron opacity peak and
density inversion at $78\rsun$.  The density inversion region is,
however, convectively unstable and starts to mix at time $9.4t_0$ as
shown in the left panel of Figure \ref{StarMidRho}.  At time
$18.8t_0$, the whole envelope is turbulent (right panel of
Fig. \ref{StarMidRho}). This temporal evolution is very similar to the
run {\sf StarDeep}, but the properties of the turbulence are different
as we now describe. 

Figure \ref{StarMidHist} shows the radiation energy density and
density weighted vertical velocities ($V_{z,E_r},V_{z,\rho}$) as a
function of time.  After the initial density inversion is broken
around $10t_0$, $ 2.4\%$ of the mass is lost through the open top
boundary. The averaged vertical velocity also reaches the peak at the
same time. The whole envelope oscillates with a period of $4t_0$.
The vertical velocities can reach $10\%$ of the typical radiation
sound speed, which is much larger than the oscillation amplitude in
{\sf StarDeep} (Figure \ref{StarDeepHist}).  When the envelope
expands, it is accelerated by the radiation without the density
inversion. But because of the sensitive dependence of the iron opacity
on temperature and the temperature decrease with height, the radiation
acceleration becomes smaller than $g$ at some height and the
acceleration stops. The whole envelope starts to fall back when the
vertical velocity reverses.

\begin{figure}[h!]
\begin{center}
\includegraphics[width=1.0\columnwidth]{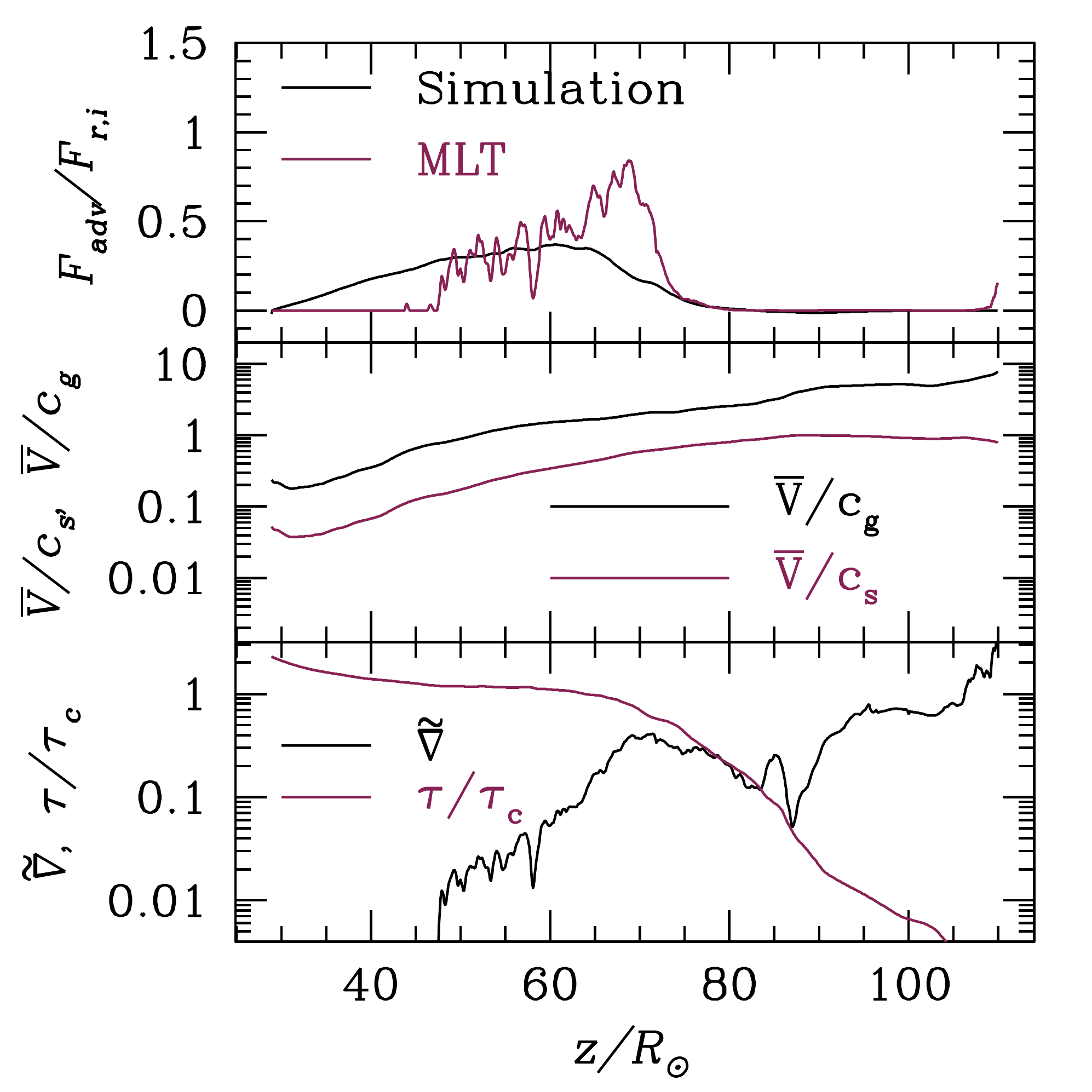}
\caption{Top: time and horizontally averaged vertical profiles of 
advection flux (black line) and convection flux 
predicted by mixing length theory (red line). Middle: the averaged 
turbulent velocity $\overline{V}$ scaled with the radiation sound speed $c_s$ (red line) 
and isothermal sound speed (black line). Bottom: vertical profile of the superadiabatic parameter 
$\tilde{\nabla}$ (black line) and the ratio between the local optical depth 
per pressure scale height $\tau$ and the critical value $\tau_c$ (red line). 
These profiles are for the run {\sf StarMid} and 
the time average is done between $17.6t_0$ and $29.2t_0$.%
}
\label{StarMidConvection}
\end{center}
\end{figure}

\begin{figure*}[htp]
\begin{center}
\includegraphics[width=1.0\columnwidth]{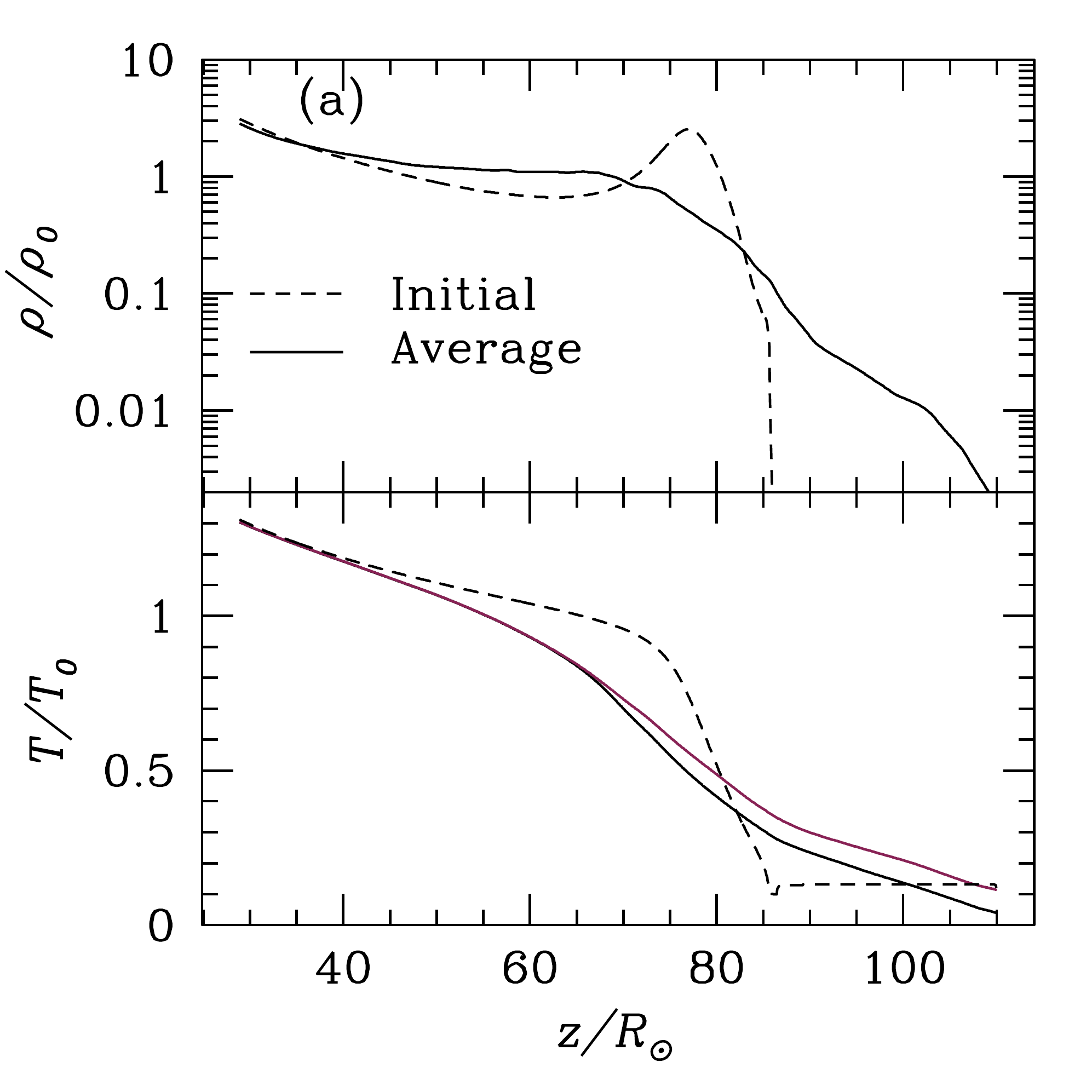}
\includegraphics[width=1.0\columnwidth]{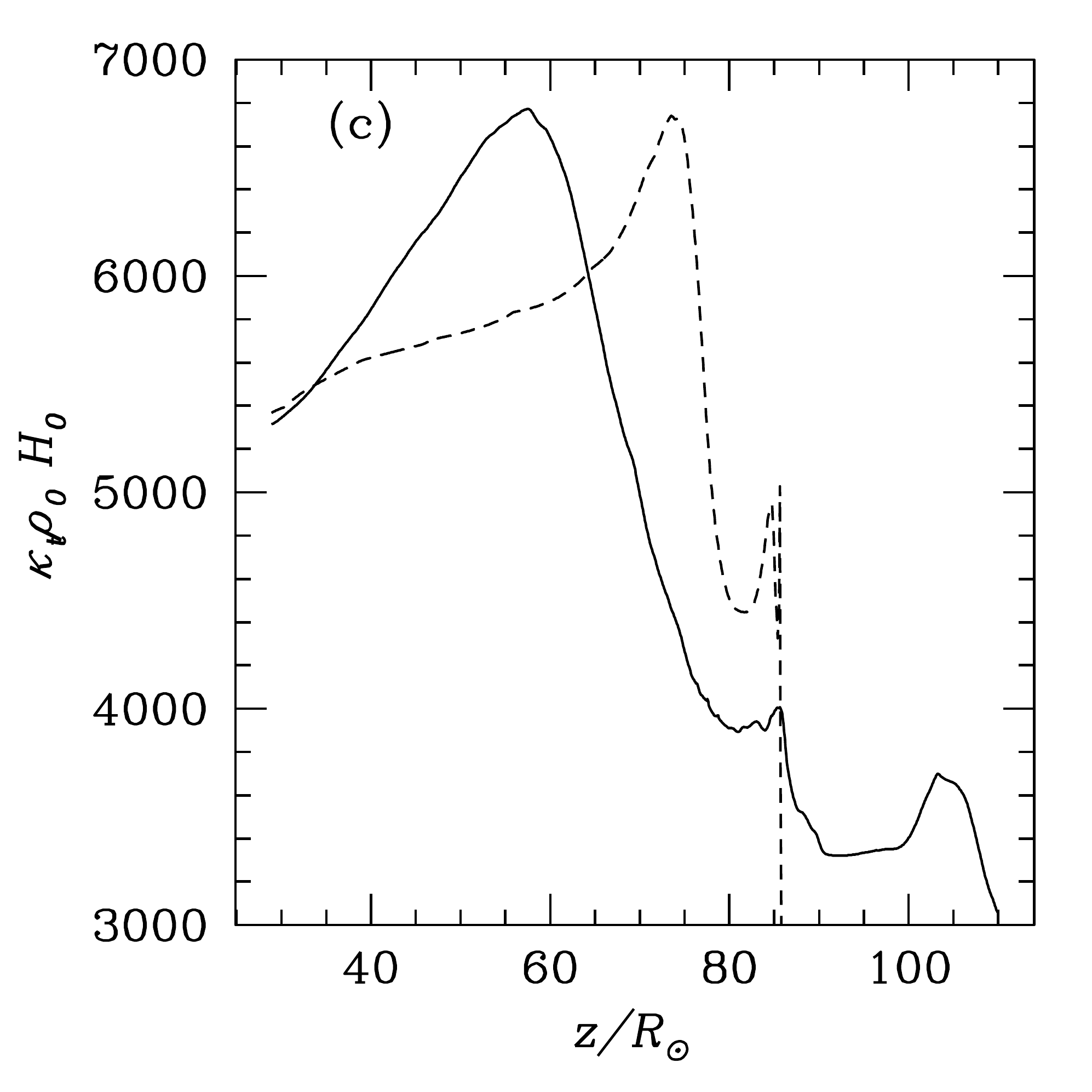}\\
\includegraphics[width=1.0\columnwidth]{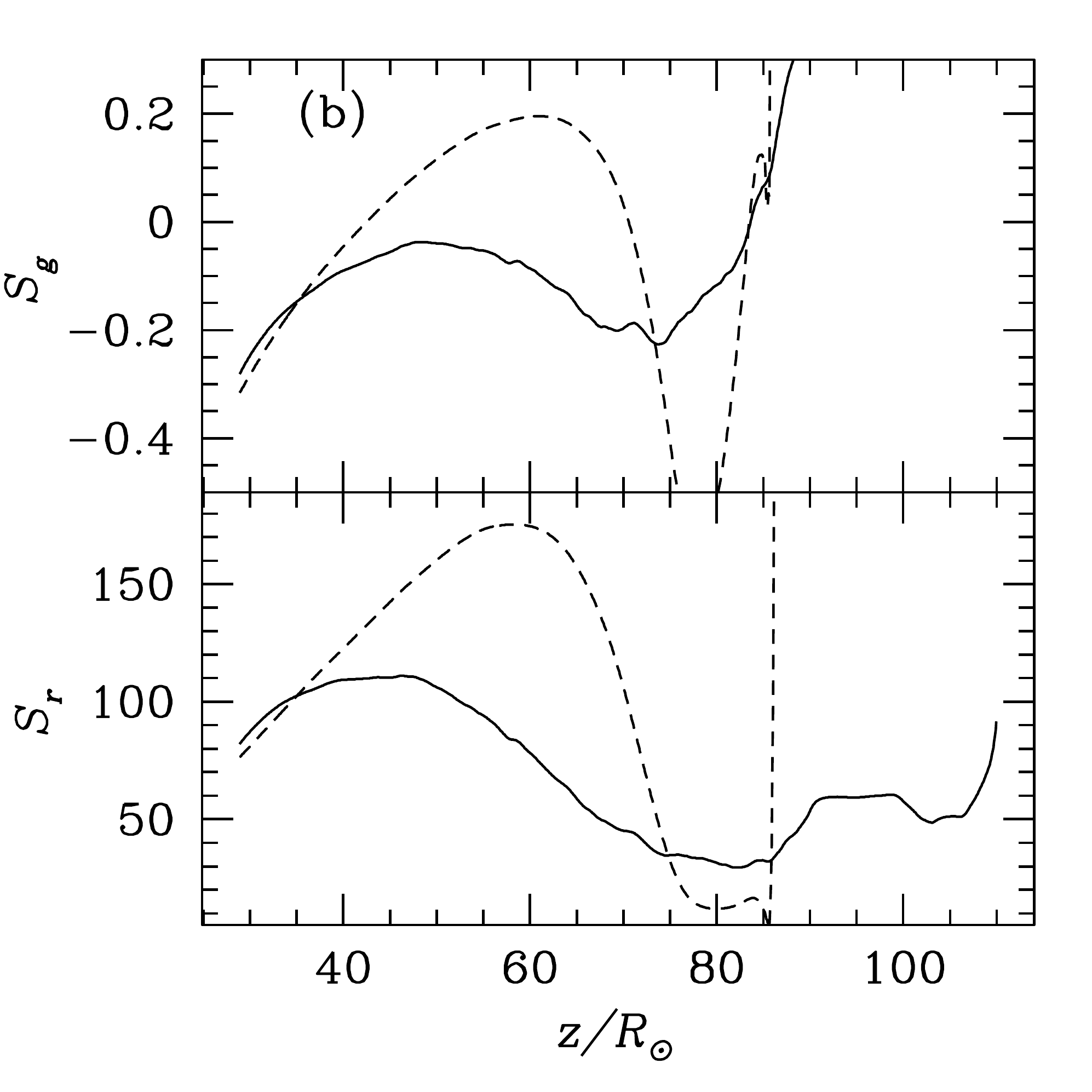}
\includegraphics[width=1.0\columnwidth]{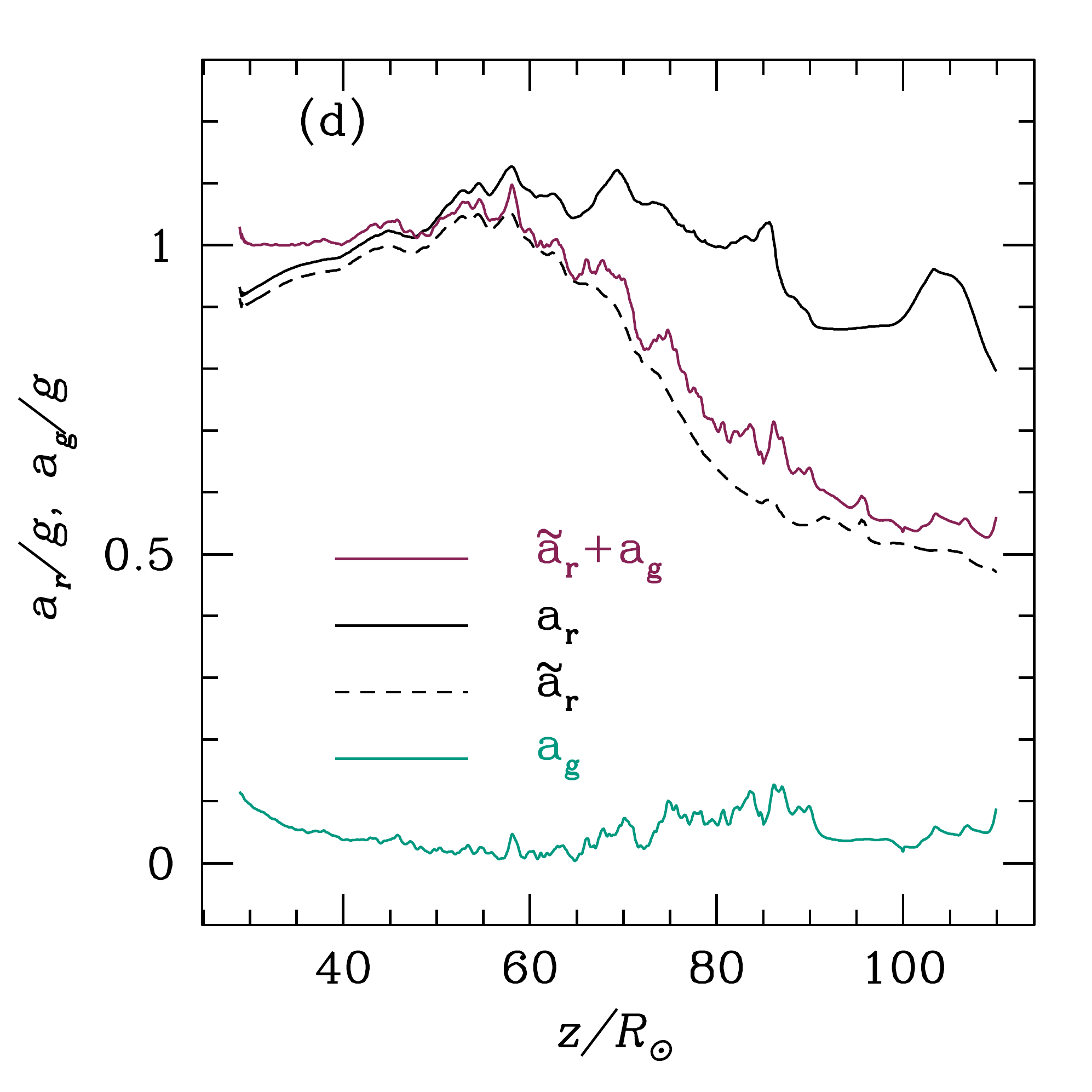}
\caption{The time and horizontally averaged vertical profiles of density (top panel of a), temperature 
(bottom panel of a, solid black line for gas temperature and red line for radiation temperature), 
entropy (panel b), opacity (panel c) as well as accelerations (panel d) for the 
run {\sf StarMid}. As in Figure \ref{StarDeepProfile}, the dashed lines in panels (a), (b), (c) 
are initial conditions while other lines are time averaged profiles. The initial condition 
above $85R_{\odot}$ is the  region above the photosphere where we set it to be isothermal. 
}
\label{StarMidProfile}
\end{center}
\end{figure*}

\subsubsection{Time Averaged Vertical Structure and Turbulent Properties}

\begin{figure}[h!]
\begin{center}
\includegraphics[width=1.0\columnwidth]{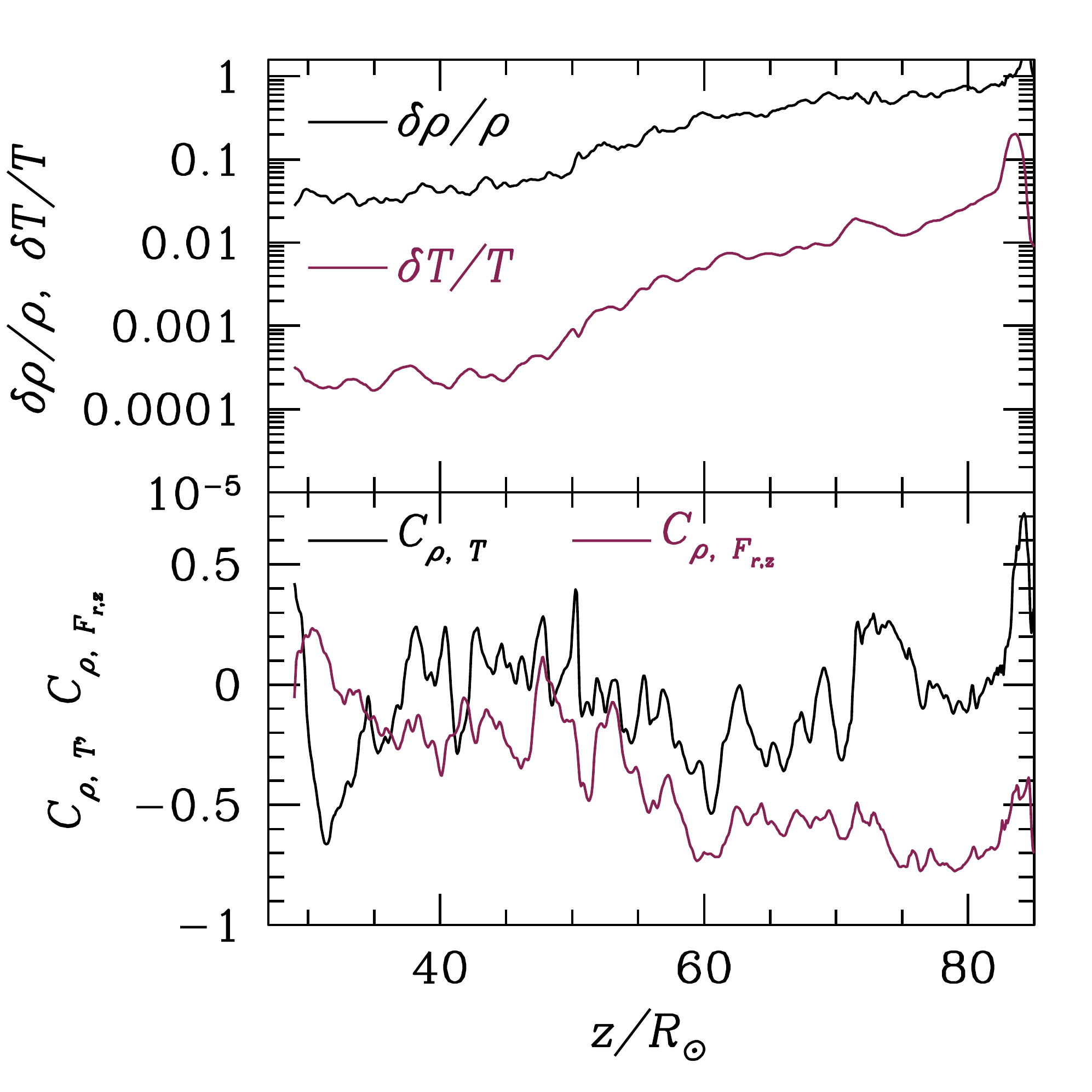}
\caption{Top: vertical profiles of density and temperature standard deviations scaled with the 
horizontally averaged mean values. Bottom: vertical profiles of cross correlation coefficients between 
$\rho, T$ and $\rho, F_{r,z}$. These profiles are for the run {\sf StarMid} 
at time $18.8t_0$.%
}
\label{StarMidsigma}
\end{center}
\end{figure}

Figure \ref{StarMidConvection} (solid line) shows that there is
significant turbulent transport of energy below $ 65 \rsun$, but
the turbulent transport falls off rapidly at larger heights.  At
about the same location $ 65 \rsun$, the averaged turbulent
velocity $\overline{V}$ reaches the isothermal sound speed (middle
panel of Fig. \ref{StarMidConvection}), but is still smaller than the
radiation sound speed.  

Figure \ref{StarMidProfile} shows the time and horizontally averaged
vertical profiles of density, temperature, opacity, entropy and
accelerations for this run.  The iron opacity peak moves inward somewhat
in radius, but the density inversion in the initial condition is
largely gone, replaced with an extended region below $65\rsun$
over which the density changes very slowly with height.  Gas and radiation
are thermally coupled in this region. However, above $65\rsun$,
the density drops quickly with height and the gas temperature starts
to deviate from the radiation temperature. Here
the local optical depth per pressure scale height
$\tau$ becomes smaller than the critical value $\tau_c$ and the photon
diffusion time becomes smaller than the local dynamic time. This
transition to $\tau \lesssim \tau_c$ is why the advection flux drops
significantly in this region (Fig. \ref{StarMidConvection}).  

The top panel of Figure \ref{StarMidConvection} shows the convective
flux calculated based on the time averaged vertical profiles and
mixing length theory (eq. \ref{eq:mlt}) with $\alpha=0.45$, which is
again close to the ratio between the correlation length $l_{\rho}$ and
scale height $H_0$.  Below $60\rsun$ where $\tau > \tau_c$, the time
averaged advection flux is comparable to the mixing length theory
convection flux, which is consistent with Figure
\ref{StarDeepConvection} for the run {\sf StarDeep}. However, above
$65\rsun$ the advection flux drops significantly in the
simulations, but equation (\ref{eq:mlt}) still predicts a very large
convection flux because of the decreasing radiation entropy (see panel b
of Fig \ref{StarMidProfile}).  This demonstrates that for
$\tau \lesssim \tau_c$, convection becomes inefficient and equation
(\ref{eq:mlt}) significantly overestimates the amount of advection flux
with a constant $\alpha$.
 
The vertical profile of $\tau/\tau_c$, where 
$\tau=\rho \kappa_t H_0$ is the local optical depth per pressure scale 
height, shown in the bottom panel of Figure \ref{StarMidConvection} 
confirms that the transition to inefficient convection 
happens when $\tau/\tau_c<1$. 
It also happens roughly when the turbulent velocity $\overline{V}$ becomes
larger than the isothermal sound speed (the middle panel of Figure
\ref{StarMidConvection}), and the superadiabatic parameter
$\tilde{\nabla}$ increases significantly to $\sim 0.4$ (bottom panel
of Figure \ref{StarMidConvection}). Notice that throughout the whole
envelope, $\tilde{\nabla}$ is comparable to $\overline{V}/c_s$ as in
Figure \ref{StarDeepConvection} for the case {\sf StarDeep}.

Panel (d) of Figure \ref{StarMidProfile} shows that the volume
averaged radiation acceleration $a_r$ is larger than the gravitational
acceleration between $50\rsun$ and $80\rsun$, which is also the region
where radiation entropy decreases with height.  It is thus perhaps
surprising that when convection becomes inefficient above
$65\rsun$ and there is little turbulent transport of energy, the
envelope does not develop a density inversion again. 
This is because 
the density weighted radiation acceleration $\tilde{a}_r$
(eq . \ref {eq:artilde}, shown as the dashed black line in panel (d) of
Figure \ref{StarMidProfile}) is significantly smaller than the volume
averaged radiation acceleration $a_r$ above $65\rsun$.  The sum
$\tilde{a}_r+a_g$ is comparable to $g$ at the iron opacity peak around
$60\rsun$. Beyond $65\rsun$, it drops below $g$ significantly.

Figure \ref{StarMidsigma} clarifies the origin of the reduced density
weighted radiation acceleration.  In particular, it shows that density
fluctuations reduce the effective radiation acceleration above
$65\rsun$ because there is an anti-correlation between
fluctuations of density and radiation flux.  The correlation
coefficient $C_{\rho,F_{r,z}}$ in Figure \ref{StarMidsigma} shows this
explicitly and is always negative. However, the nature of the
anti-correlations changes at different heights. Below $65\rsun$, when
$\tau>\tau_c$, the situation is similar to the run {\sf StarDeep}, in
that buoyancy causes the anti-correlation between $\rho$ and
$F_{r,z}$, as $F_{r,z}$ is dominated by the advection flux. Above
$65\rsun$, however, when $\tau \lesssim \tau_c$, the situation
changes.  Convection becomes inefficient and $F_{r,z}$ is dominated by
the diffusive flux (the advection flux drops as shown in Figure
\ref{StarMidConvection}). The anti-correlation between $\rho$ and
$F_{r,z}$ above $65\rsun$ is thus not caused by
buoyancy. Instead, it is because the ``porosity" of the envelope
created by the traveling strong shocks allows radiation 
to propagate through low density
regions \citep{Shaviv1998}. This reduces the effective radiation acceleration
significantly (see panel (d) of Figure \ref{StarMidProfile}).
Note also that in {\sf StarMid}, the standard 
deviations of density and temperature (scaled with the horizontally 
averaged quantities) are significantly larger 
than the corresponding values in the run {\sf StarDeep} (Figure \ref{StarDeepsigma}).
\subsection{The run {\sf StarTop} when $\tau_0\ll\tau_c$}
This is the regime when gas and photons are loosely coupled through
the whole iron opacity peak region.  Convection will be inefficient as in the
top region of {\sf StarMid}.  The photosphere is now closer to the
region with the iron opacity peak and accurate radiation transfer is
critical. For this run, we use a density floor of $10^{-10}\rho_0$ to
avoid numerical difficulties.

\begin{figure}[h!]
\begin{center}
\includegraphics[width=1.0\columnwidth]{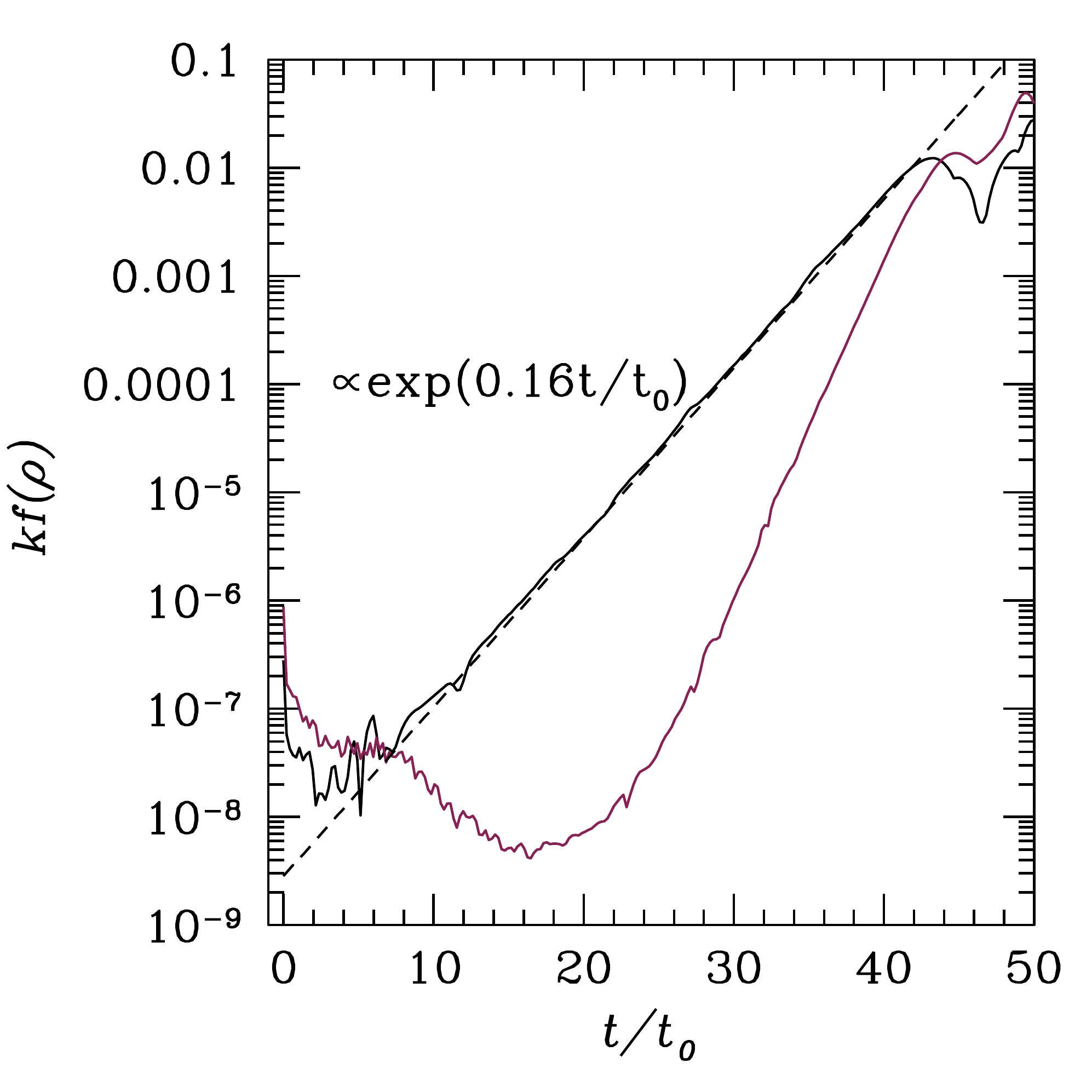}
\caption{Time evolution of the density power spectrum $kf(\rho)$ for the run {\sf StarTop} 
at height $z=13.7R_{\odot}$ during the first $50t_0$. The solid black and red lines 
are for the modes with wavelength $H_0$ and $0.26H_0$ respectively. 
The long wavelength mode has an e-folding time $6.25t_0$ as indicated by 
the dashed black line. 
}
\label{StarTopLinear}
\end{center}
\end{figure}

\begin{figure}[htp]
\begin{center}
\includegraphics[width=1.0\columnwidth]{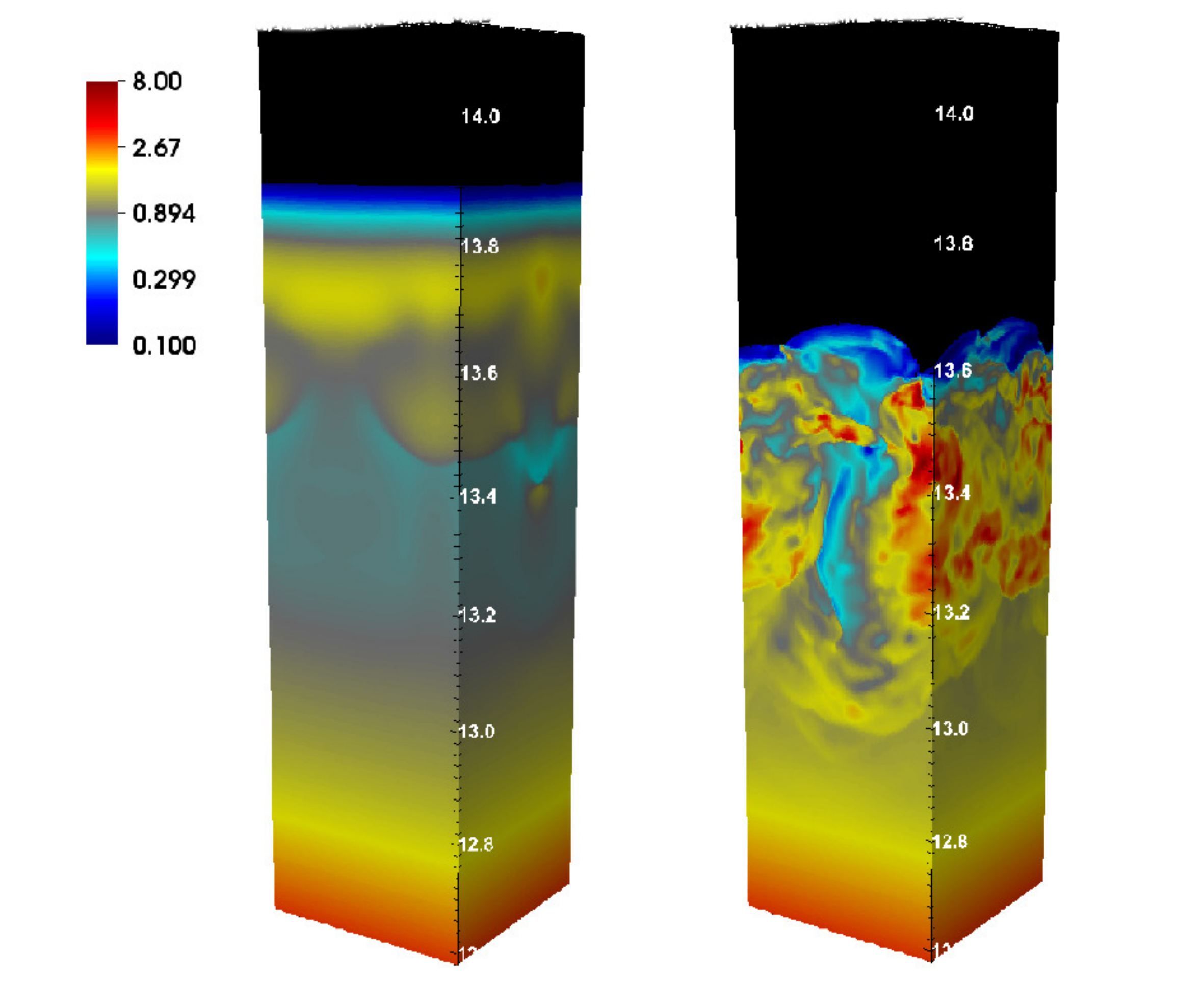}
\caption{Snapshots of density for the run {\sf StarTop} at time $49.1t_0$ (left) and 
$138.9t_0$ (right). %
}
\label{StarTopRho}
\end{center}
\end{figure}

\begin{figure}[h!]
\begin{center}
\includegraphics[width=1.0\columnwidth]{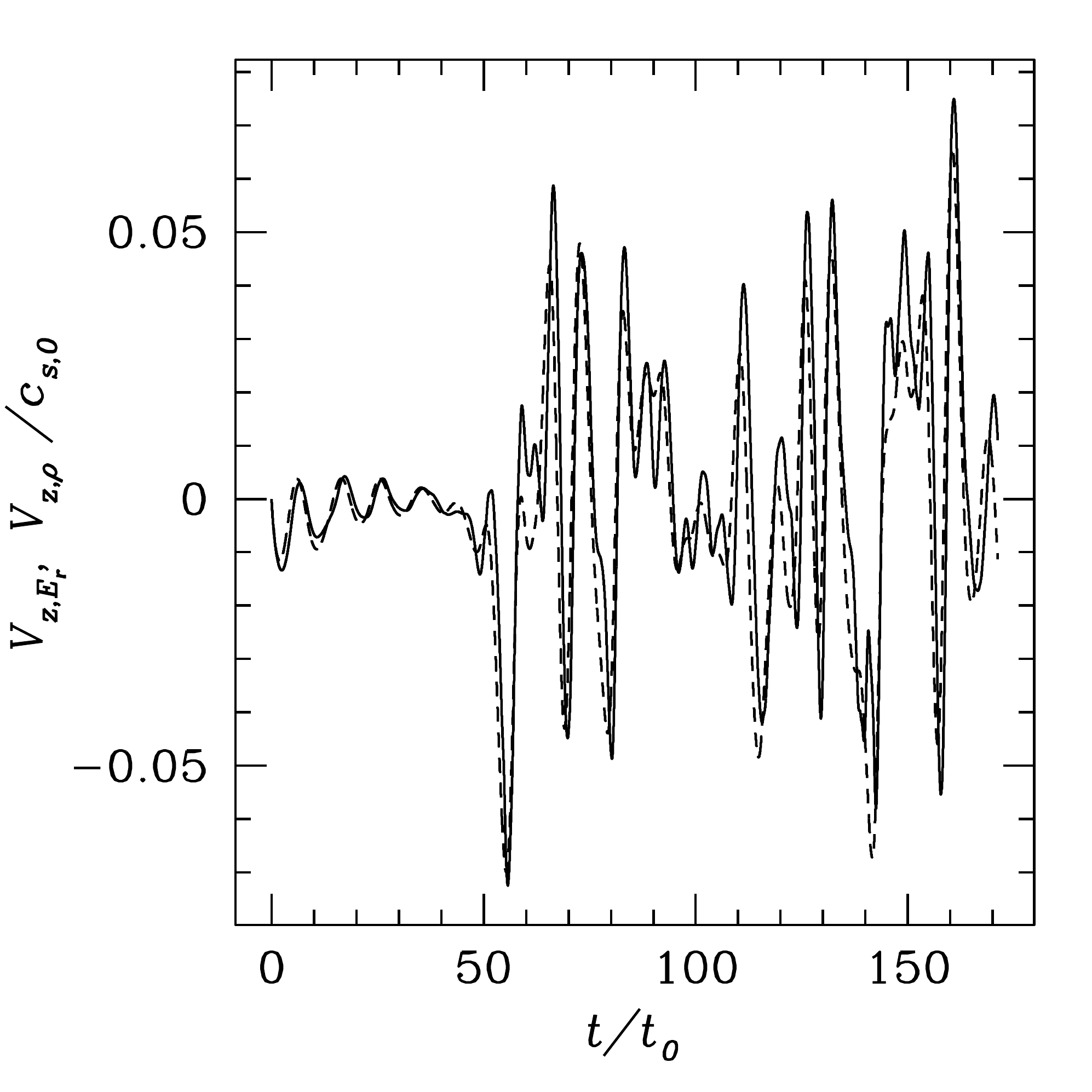}
\caption{History of the volume averaged vertical velocity $V_{z,E_r}$ (solid line at bottom) 
and $V_{z,\rho}$ (dashed line at bottom)  
for the run {\sf StarTop}. The volume average is done for the whole simulation box. }
\label{StarTopHist}
\end{center}
\end{figure}

\subsubsection{Simulation History}

\begin{figure}[h!]
\begin{center}
\includegraphics[width=1.0\columnwidth]{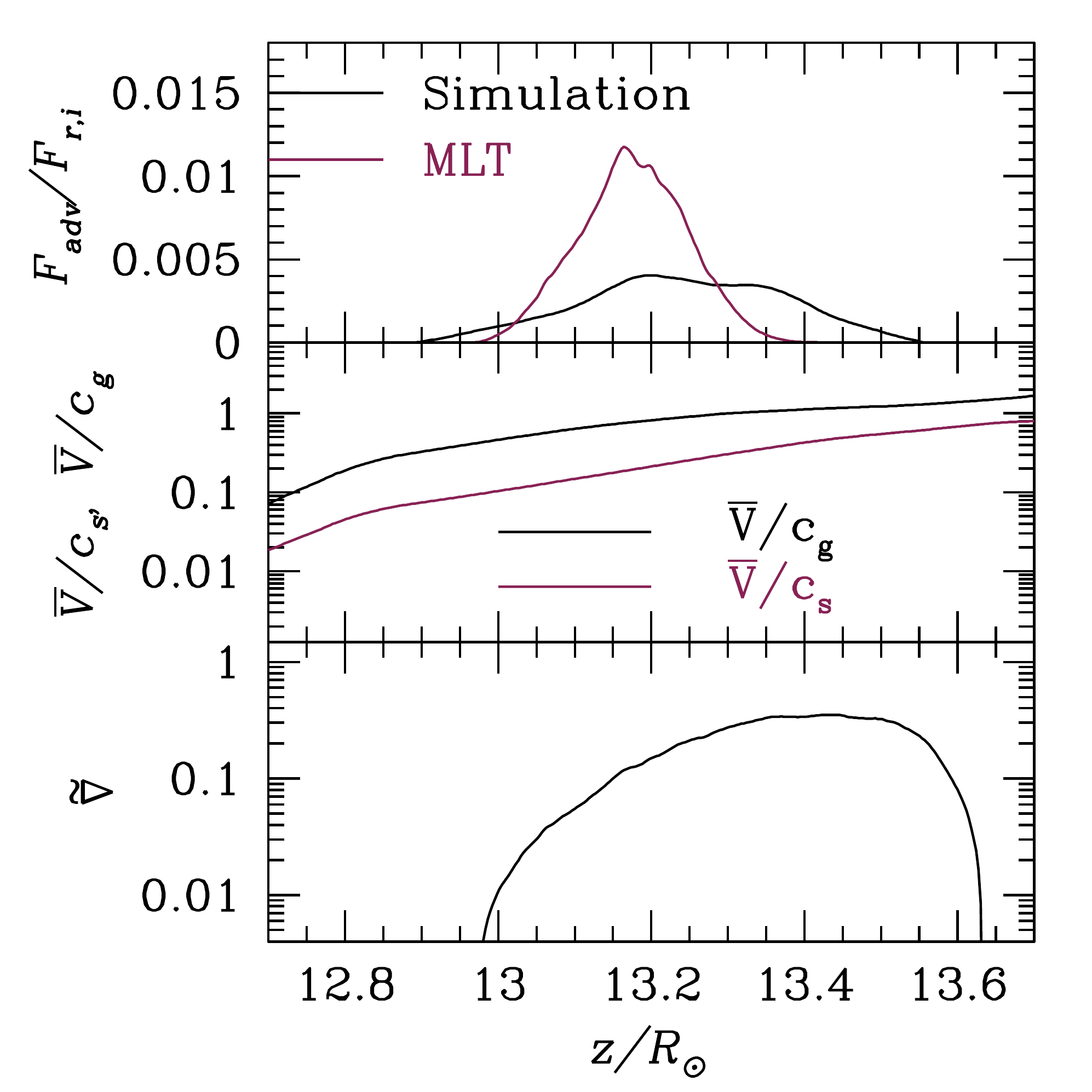}
\caption{Top: time and horizontally averaged vertical profiles of 
advection flux (black line), 
and convection flux according to the non-adiabatic MLT with $\alpha=0.4$ (equation \ref{eq:mlt})
as given by \cite{Henyey:1965} (red line). 
Middle: the averaged 
turbulent velocity $\overline{V}$ scaled with the radiation sound speed $c_s$ (red line) 
and isothermal sound speed (black line). Bottom: vertical profile of superadiabatic parameter 
$\tilde{\nabla}$. The time average is done between $55.6t_0$ and $171.0t_0$. These profiles 
are for the run {\sf StarTop}.
}
\label{StarTopConvection}
\end{center}
\end{figure}

The iron opacity peak and density inversion are located at
$13.7\rsun$ initially. This is again convectively unstable and the
initial structure is completely destroyed by $50t_0$. Just as we
found in the simulation {\sf StarDeep}, this time scale is
significantly longer than the buoyancy time scale $1/|N_r|$, which is
$\approx t_0$ at the iron opacity peak. We therefore again explored the
linear growth phase for this run by calculating the density power
spectrum at the location of the initial density inversion.  The time
evolutions of the power $kf(\rho)$ in two modes with horizontal
wavelengths $H_0$ and $0.26H_0$ are shown in Figure
\ref{StarTopLinear}.  As in {\sf StarDeep} (Fig.
\ref{StarDeepLinear}), the long wavelength mode grows exponentially
from close to the beginning, with an e-folding time of
$6.25t_0$. However, in contrast to the {\sf StarDeep} case, the short
wavelength mode is damped initially and only starts to grow quickly
after $20t_0$ when the e-folding time is $3.33t_0$.  This later
rapid growth is almost certainly due to a nonlinear cascade.

We have examined the behavior of other horizontal wavelengths as well,
and found that all the modes with horizontal wavelengths larger than
$\approx 0.4H_0$ have similar growth rates after $12t_0$, while the modes with
horizontal wavelengths smaller than $\approx 0.3H_0$ are damped until the
nonlinear cascade sets in. Based on the analysis of
\cite{BlaesSocrates2003}, we estimate that modes with wavelengths less
than about $2.5 H_0$ are in the regime where radiative diffusion
reduces the growth rate of the convective instability.   This is
larger than the size of the box, so that all convective modes in the
simulation are heavily modified by radiative diffusion (consistent
with the fact that $\tau_0 \lesssim \tau_c$).
From equation (59) of \cite{BlaesSocrates2003}, we find that the long
wavelength mode in Figure \ref{StarTopLinear} should have a growth
rate of $\simeq0.17/t_0$, assuming a vertical wavelength that is
comparable to the horizontal wavelength (which is also comparable to
the vertical width of the density inversion).  This agrees very well
with the measured growth rate.  However, shorter wavelength modes
should have smaller growth rates $\propto k^{-2}$, in contrast to the
behavior measured in the simulations.  It is not clear what is causing
this discrepancy, although we suspect that the damping of modes 
with wavelength $\lesssim 0.3H_0$ is due to the
numerical damping rate being larger than the very 
small growth rate for these modes. 
Interestingly, traveling acoustic waves should also be
unstable according to equation (62) of \cite{BlaesSocrates2003}, with
comparable growth rates to the buoyancy instability (such waves
should not be unstable in the {\sf StarDeep} simulation).
Vertically propagating acoustic waves (with infinite horizontal wavelength)
should grow fastest, but these would not necessarily show up as growing
perturbations at a fixed height.  We have not further investigated this
possibility, as it is clear that the unstable buoyancy
modes dominate the evolution observed in the simulation.

Figure \ref{StarTopRho} shows snapshots of the density structure of
the envelope at time $49.1t_0$ and $138.9t_0$ while the histories of
the vertical velocities $V_{z,E_r}, V_{z,\rho}$ are shown in Figure
\ref{StarTopHist}.  The whole envelope becomes turbulent after the
initial density inversion is destroyed around $50t_0$.  Unlike the
simulations {\sf StarDeep} and {\sf StarMid}, the turbulent envelope
is not a relatively quiescent convective structure, but instead shows
violent, irregular, large amplitude oscillations.  The oscillation
period varies from $\sim5$ to $10t_0$. The amplitude of $V_{z,E_r}$ is
$0.05c_{s,0}$, which is already $\approx18\%$ of the isothermal sound
speed.  At time $138.9t_0$ when $V_{z,E_r}$ reaches a minimum, the
right panel of Figure \ref{StarTopRho} shows that there is even a low
density hole extending over one pressure scale height $H_0$.  The mass
loss through the open top boundary is negligible for this run.

\begin{figure*}[htp]
\begin{center}
\includegraphics[width=1.0\columnwidth]{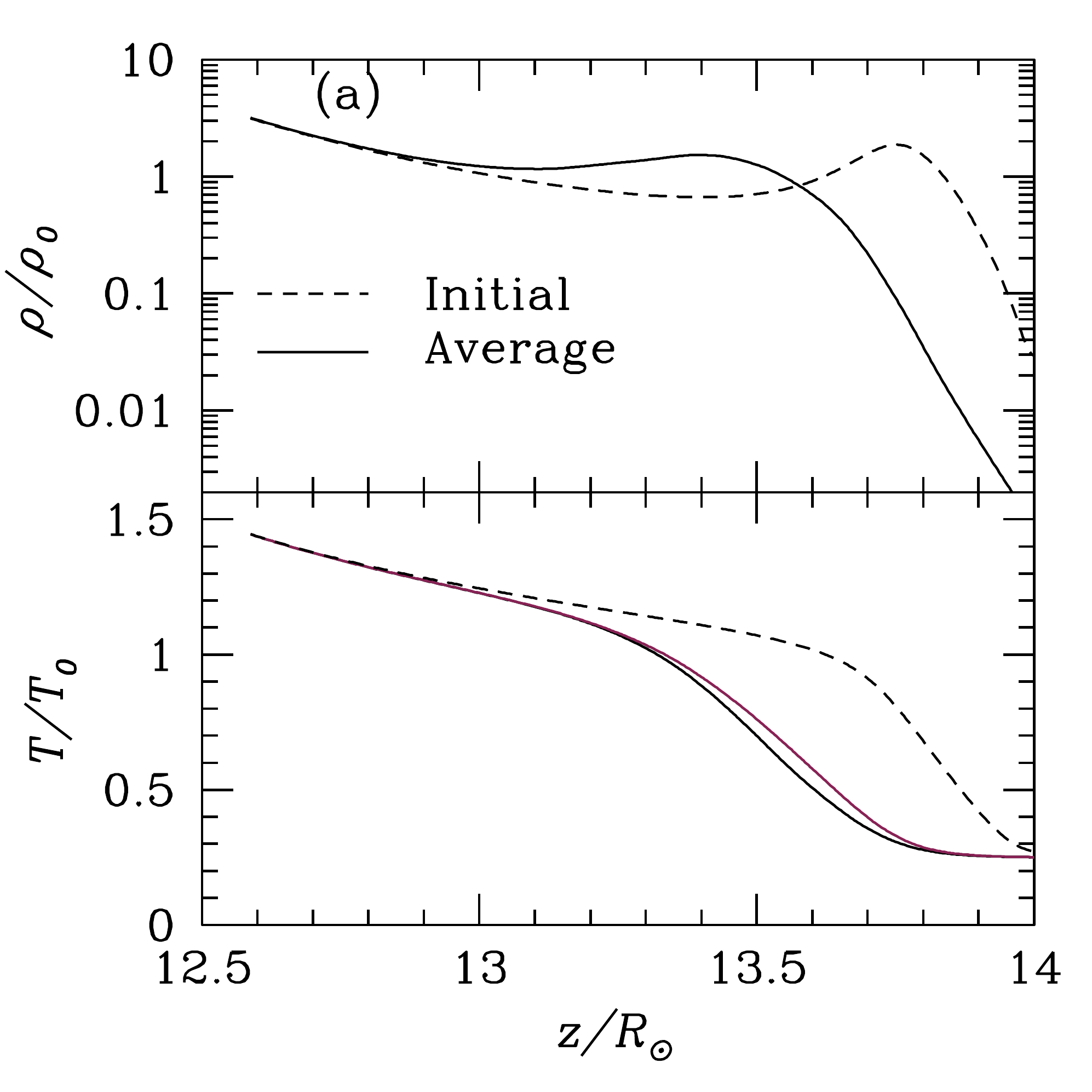}
\includegraphics[width=1.0\columnwidth]{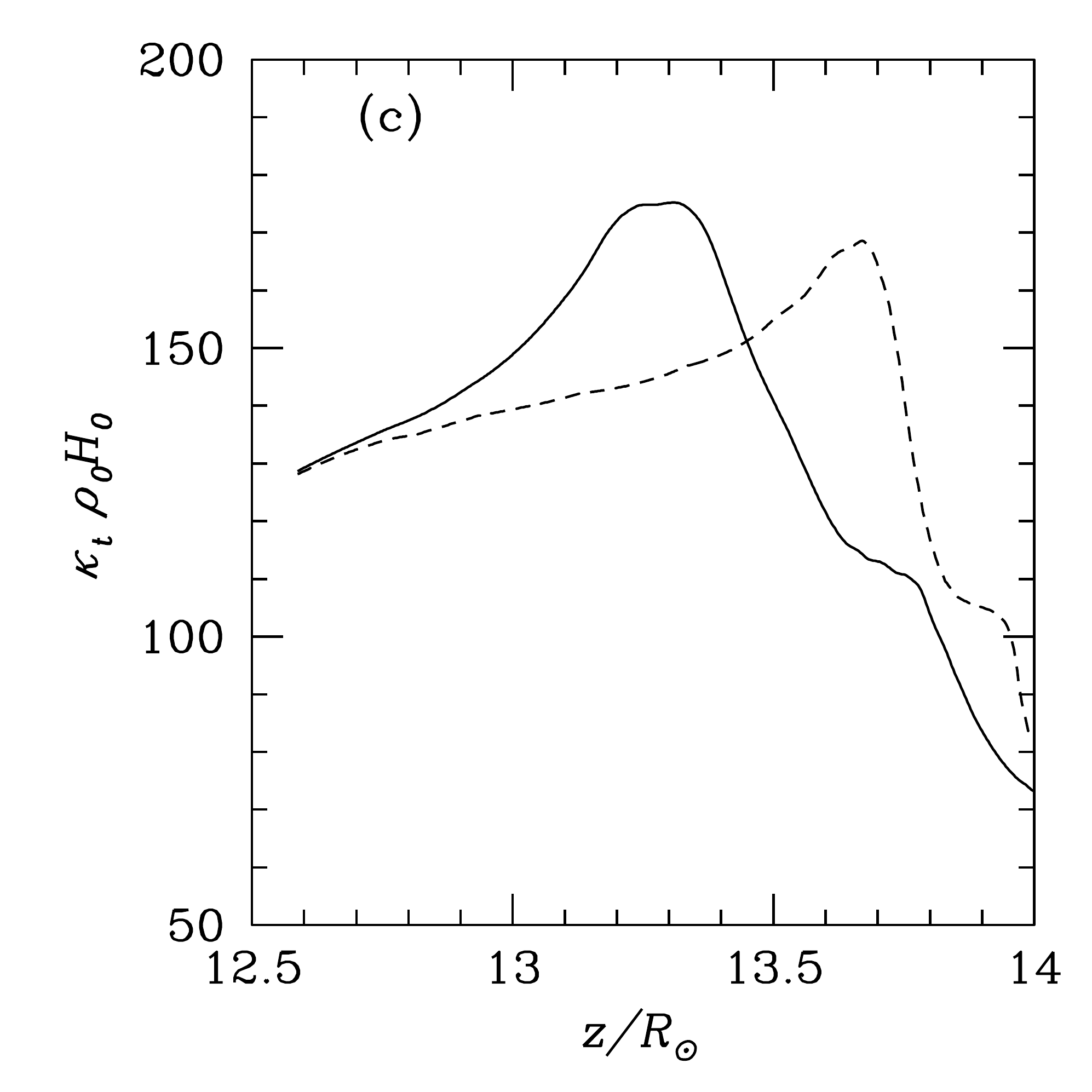}\\
\includegraphics[width=1.0\columnwidth]{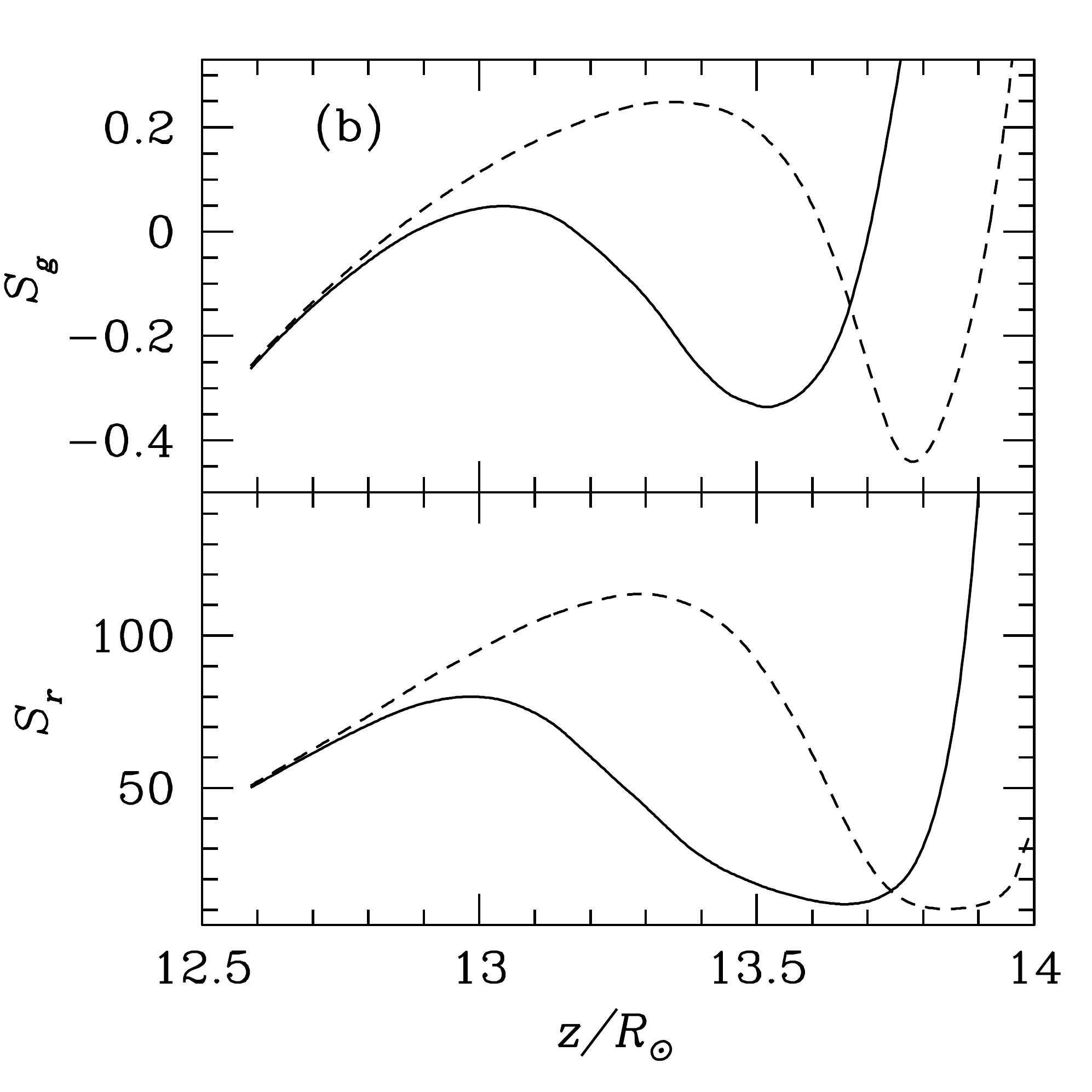}
\includegraphics[width=1.0\columnwidth]{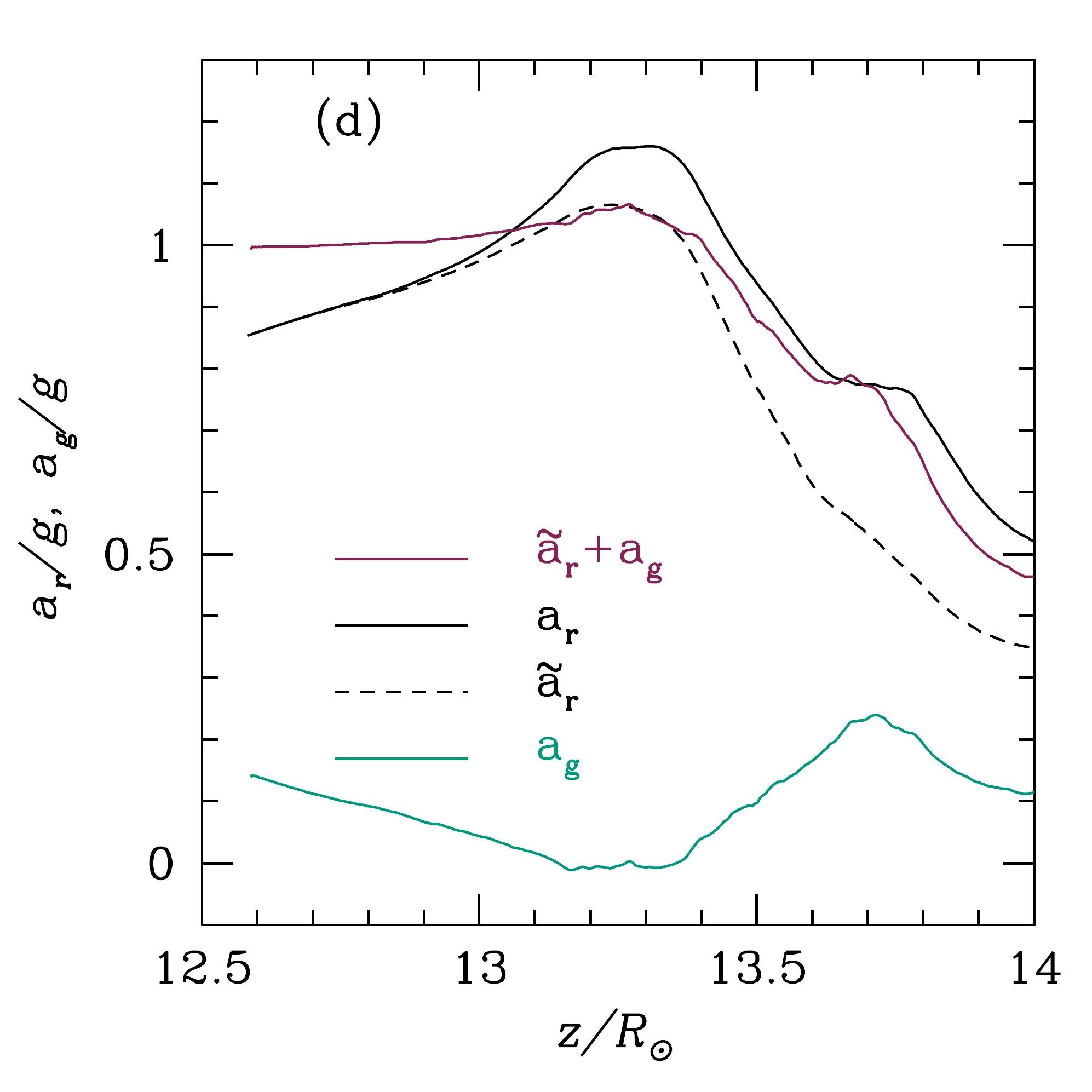}
\caption{The time and horizontally averaged vertical profiles of density (top panel of a), 
temperature (bottom panel of a, black and red lines for gas and radiation temperature), 
entropy (panel b), opacity (panel c) as well as accelerations (panel d) for the 
run {\sf StarTop}. The dashed lines in (a), (b) and (c) are initial conditions while other lines 
are time averaged profiles between $55.6t_0$ and $171.0t_0$.%
}
\label{StarTopProfile}
\end{center}
\end{figure*}

According to equation (\ref{eq:tc}), the typical thermal time scale is
$t_c\sim0.09t_0$ for this run.  However, for this $\tau_0 \ll \tau_c$
regime, $t_c$ is no longer the most relevant time scale
for some of the dynamics. Instead, the radiation flux behaves like a
roughly constant force reducing the effective gravity
\citep[][]{Jiangetal2013a}.  The more relevant time scale is the
effective dynamic time scale
\begin{eqnarray}
\tilde{t}_0\equiv \frac{c_{g,0}}{|\tilde{g}|},
\end{eqnarray}
where $c_{g,0}$ is the isothermal sound speed and
 $\tilde{g}\equiv g-\tilde{a}_r$ is the effective acceleration due to
gravity and the radiation force.  For an effective acceleration
$\tilde{g}=-0.06g$ according to the $\tilde{a}_r$ shown in panel (d) of
Figure \ref{StarTopProfile} (discussed below), the effective dynamic
time scale is $\tilde{t}_g=4.6t_0$, which is consistent with the
oscillation period of the envelope in Figure \ref{StarTopHist}.  Note
that the effective scale height is then
$\tilde{H}_0\equiv c_{g,0}^2/|\tilde{g}|=1.26H_0$, which happens to be
similar to $H_0$.


\subsubsection{Time Averaged Vertical Structures}

Figures \ref{StarTopConvection} and \ref{StarTopProfile} show vertical
profiles of various time and horizontally averaged quantities in the
nonlinear state after $55.6t_0$.  Unlike the previous two runs {\sf
  StarDeep} and {\sf StarMid}, the averaged advection flux is less
than $0.5\%$ of $F_{r,i}$ in the whole envelope as shown in the top
panel of Figure \ref{StarTopConvection}.  This is because
$\tau_0\ll \tau_c$ everywhere and gas is not able to trap photons. The
averaged radiation flux $F_{r,0z}$ is almost a constant vertically
except very near the top of the domain; the radiation flux is nearly
constant in time as well, with $\lesssim 8\%$ variation during the
whole simulation.  

Because the radiation flux dominates over the advection flux, the full
super-Eddington flux contributes to the radiation acceleration of the
gas.  However, because of the large density fluctuations, most photons
go through regions that have below average density.  This reduces the
density weighted radiation acceleration $\tilde{a}_r$ (eq. \ref
{eq:artilde}), as shown in panel (d) of Figure
\ref{StarTopProfile}. However, unlike the assumptions made in previous
models \citep[][]{Shaviv1998}, the ``porosity" of the envelope due to
convection cannot reduce the effective radiation acceleration to a
value below $g$. At $z\approx13.3\rsun$ where the iron opacity peak is
located now, $\tilde{a}_r$ is still about $6\%$ larger than $g$, while
$a_r$ is about $10\%$ larger than $g$. This is why the average density
profile still shows a mild inversion near the iron opacity peak, as
shown in panel (a) of Figure \ref{StarTopProfile}.  We stress that
although the density inversion remains, the resulting structure is
nonetheless very different from the hydrostatic density inversion in
the initial condition.  The high density region near the iron opacity
peak is continuously reformed and then mixed, triggering violent
oscillations, shocks, and large density fluctuations in the stellar
envelope.  Physically, in the regime $\tau_0\ll \tau_c$, photon
diffusion is so rapid that radiation pressure does not respond to the
density and velocity fluctuations. As the turbulent velocity becomes
larger than the isothermal sound speed, shocks are formed in the
envelope (Figure \ref{StarTopRho}), which causes the large density
fluctuations. This phenomenon is also observed in radiation
magneto-hydrodynamic simulations of black hole accretion disks
\citep[][]{Turneretal2003,Jiangetal2013b}.

Figure \ref{StarTopProfile} shows that the radiation and gas entropies
$S_r$ and $S_g$ decrease rapidly with height around the iron opacity
peak. If we still adopt $l_{\rho}/H_0\approx 0.4$ for the mixing length
parameter $\alpha$, equation (\ref{eq:mlt})  
 over-predicts the flux relative to the simulations by a factor of more than  $\sim3$, 
and the location of the peak in the predicted flux also offsets from the peak of time averaged advection 
flux given by the simulations, 
highlighting that existing models of inefficient convection in the radiation pressure 
dominated regime are not accurate compared to our simulation results. 
The failure of convection to carry a significant energy flux is
intimately connected with the turbulent velocities becoming
supersonic, at least relative to the gas (isothermal) sound speed.
This is expected on theoretical grounds (\S \ref{sec:mesamodels}) and
is also consistent with the simulation results, as shown in the middle
panel of Figure \ref{StarTopConvection}.  The superadiabatic parameter
$\tilde{\nabla}$ is also comparable to $\tilde{V}/c_s$ and similar to
the top region in {\sf StarMid} (Figure \ref{StarMidConvection}).

\begin{figure}[htp]
\begin{center}
\includegraphics[width=1.0\columnwidth]{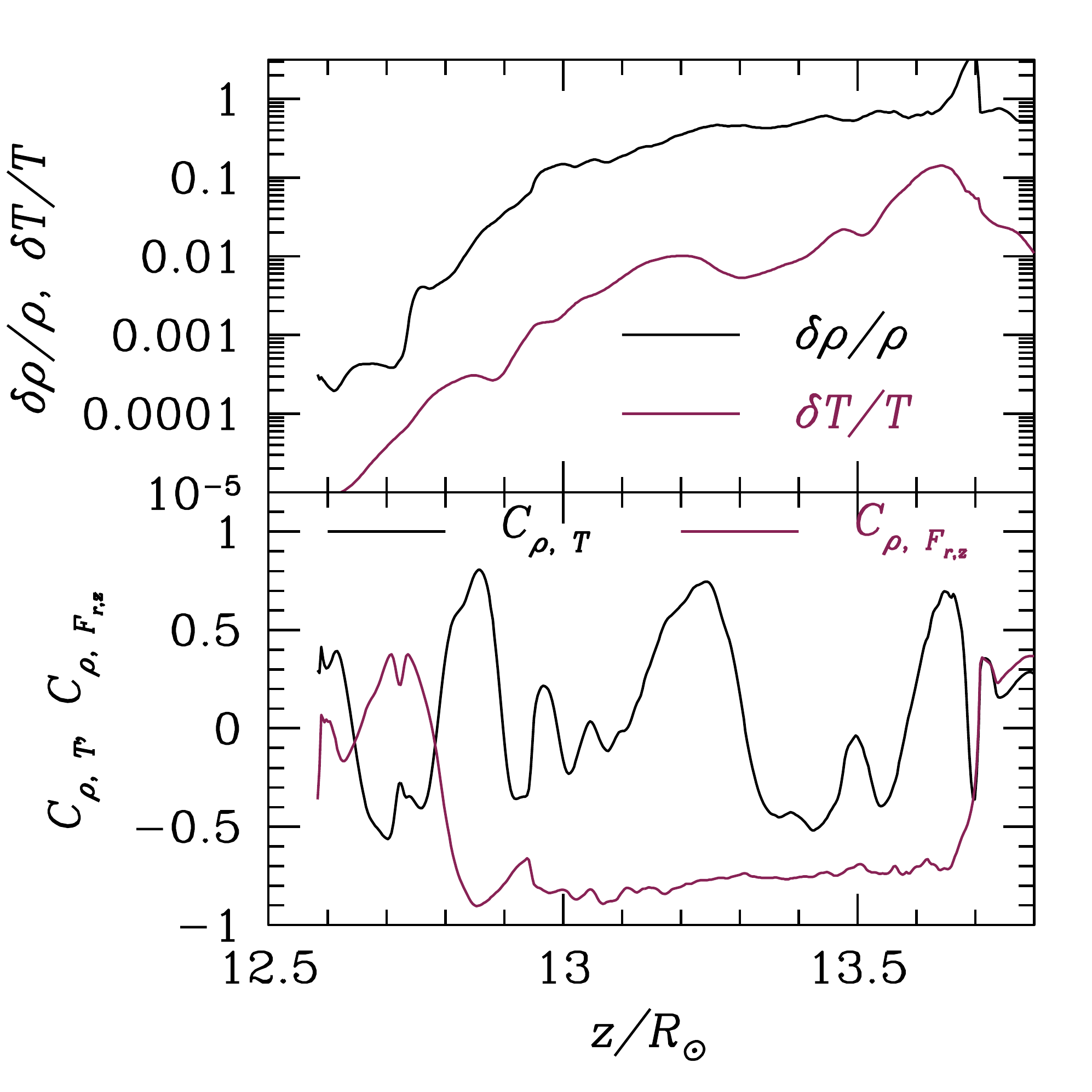}
\caption{Top: vertical profiles of density and temperature standard deviations scaled with the 
horizontally averaged mean values. Bottom: vertical profiles of cross correlation coefficients between 
$\rho, T$ and $\rho, F_{r,z}$. These profiles are for the run {\sf StarTop} 
at time $138.9t_0$.%
}
\label{StarTopsigma}
\end{center}
\end{figure}

\subsubsection{Properties of  the Turbulent Structures}

Figure \ref{StarTopsigma} quantifies the large density fluctuations
and strong anti-correlation between density and radiation flux in
simulation {\sf StarTop}.  The correlation coefficient
$C_{\rho,F_{r,z}}$ is almost $-1$ in the turbulent region between
$12.8\rsun$ and $13.6\rsun$, while $C_{\rho,T}$ only fluctuates
around $0$. This demonstrates that unlike for {\sf StarDeep}, the
anti-correlation between density and radiation flux is not due to
buoyancy in the radiation pressure dominated regime but is instead due
to the photons preferentially diffusing through low density channels.
The scaled standard deviations $\delta\rho/\rho$ and $\delta T/T$ are
also much larger than the corresponding values in {\sf StarDeep} and
{\sf StarMid}, which is consistent with the images in Figure \ref{StarTopRho}.


The typical size of the turbulent structures can also be quantified by
the correlation length $l_{\rho}$, which varies between $0.4H_0$ and
$0.5H_0$.  This is likely set primarily by geometry as in standard
mixing length arguments.  In particular, the optical depth across the
correlation length $\tau_l$ is much smaller than $\tau_c$, which means
that photon diffusion is very rapid for all of the turbulent
fluctuations and cannot be important for setting a characteristic
scale.  The correlation length $l_{\rho}$ is also much smaller than
the characteristic turnover wavelength of acoustic waves
driven unstable by the dependence of opacity on density
\citep{BlaesSocrates2003}.  Once again, it is simply the
fact that the convective turbulence is supersonic with respect to
the isothermal sound speed in the gas, and that radiation pressure
cannot respond due to the rapid diffusion, that is producing these
nonlinear structures.

\section{Discussions and Conclusions}
\label{sec:conc}

\begin{figure}[htp]
\begin{center}
\includegraphics[width=1.0\columnwidth]{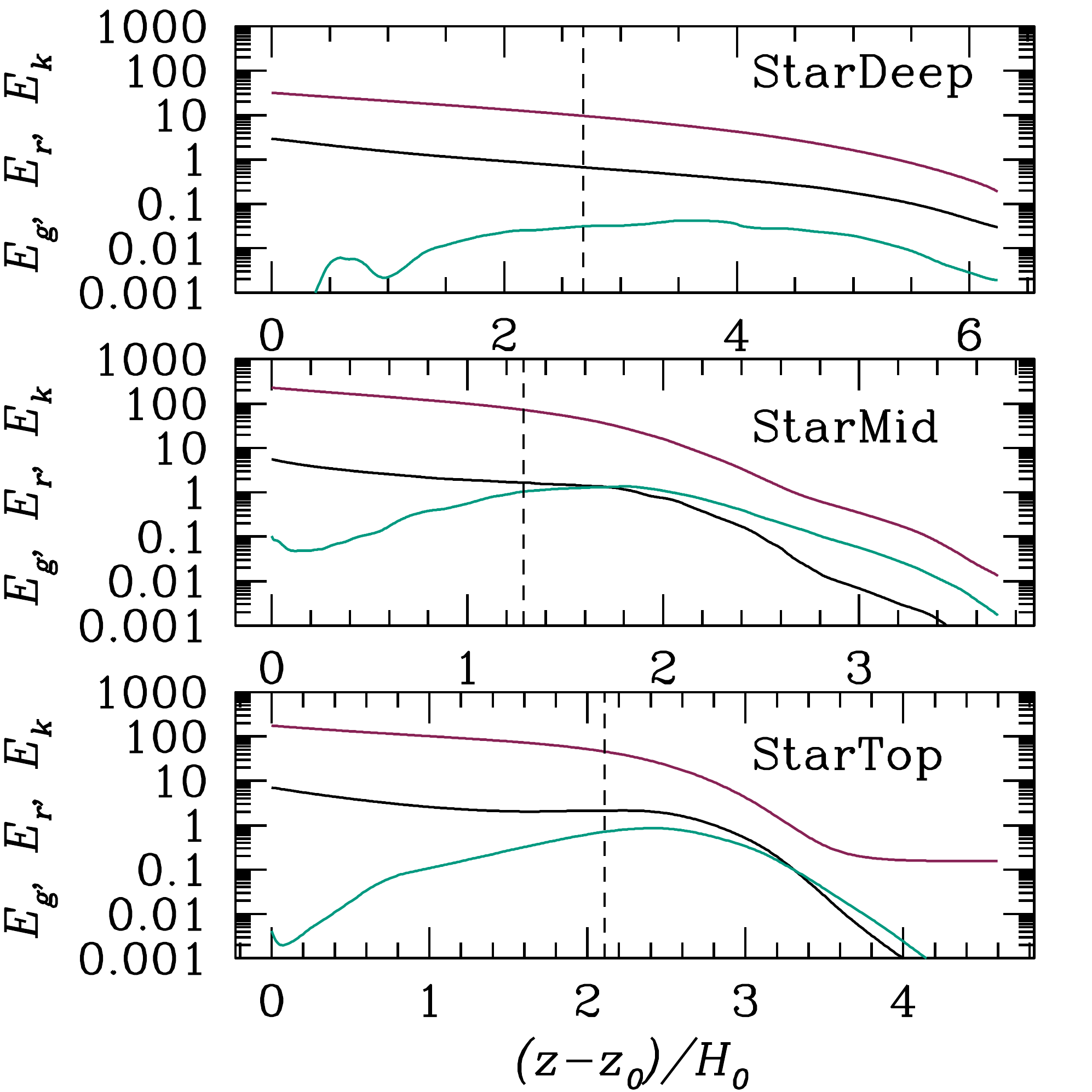}
\caption{Time and horizontally averaged vertical profiles of radiation energy density ($E_r$, red lines), 
gas internal energy density ($E_g$, solid black lines) and kinetic energy 
density ($E_k$, green lines). From top to bottom, they are for {\sf StarDeep}, 
{\sf StarMid} and {\sf StarTop} respectively. The black dashed vertical lines in each 
panel indicate the time averaged location of the iron opacity peak. The horizontal 
axis is the distance from the bottom boundary $z_0$ in units of the scale height $H_0$ 
for each run.
}
\label{CompEnergy}
\end{center}
\end{figure}

\begin{figure}[htp]
\begin{center}
\includegraphics[width=1.0\columnwidth]{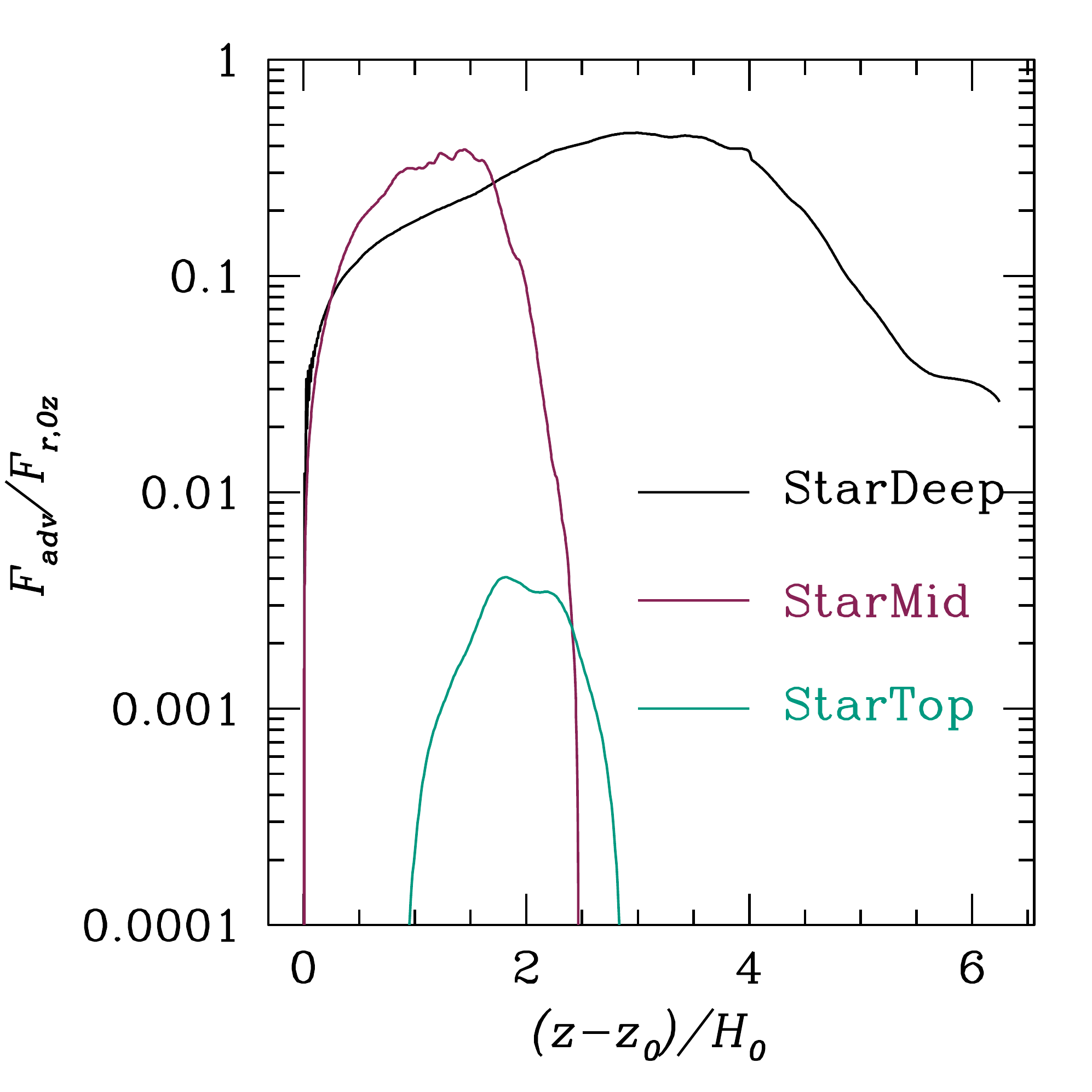}
\caption{Vertical profiles of the ratio between time averaged
  advection (convection) flux $F_{\rm adv}=v_zE_r$ and diffusive flux
  $F_{r,0z}$ for the three runs.  Convection transports significant
  energy in {\sf StarDeep} and in the deeper layers of {\sf
    StarMid}.  But in the outer layers of {\sf StarMid} and throughout
  {\sf StarTop}, photon diffusion is so rapid that it strongly
  suppresses the turbulent transport of energy.}
\label{CompFlux}
\end{center}
\end{figure}

We have presented three-dimensional (3D) radiation hydrodynamic
simulations of the radiation pressure dominated envelopes of massive
stars near the iron opacity peak.  The resulting dynamics depends
sensitively on the ratio of the photon diffusion time to the dynamical
time. This is set by the ratio of the local optical depth $\tau_0$ to
the critical optical depth $\tau_c=c/c_g$, where $c$ is the speed of
light and $c_g$ is the isothermal sound speed.  Our simulations cover
the regimes $\tau_0 \gg \tau_c$ ({\sf StarDeep}), $\tau_0 \sim \tau_c$
({\sf StarMid}), and $\tau_0 \ll \tau_c$ ({\sf StarTop}).  

The iron opacity peak at $T \approx 1.6 \times 10^5$ K leads to locally
super-Eddington fluxes in the envelopes of massive stars.  In
radiative equilibrium this produces a pronounced density inversion in
one-dimensional stellar models \citep{Jossetal1973,Paxtonetal2013}.  The
resulting density inversions are convectively unstable,
but the properties of the resulting convection zones have until now
been very uncertain when $\tau_0 \lesssim \tau_c$ and convection is
expected to be relatively inefficient.  

Figures \ref{CompEnergy} and \ref{CompFlux} compare the time and
horizontally averaged turbulent properties of our three runs, showing
the gas, radiation, and kinetic energy densities
(Fig. \ref{CompEnergy}) and the ratio of the convective and diffusive
energy fluxes, respectively.


When $\tau_0\gg \tau_c$ as in {\sf StarDeep}, convection is efficient
and the averaged radiation advection flux is consistent with the
convection flux as calculated using the MLT formalism (equation
\ref{eq:mlt}). We find that the mixing length parameter required to
match the advection flux in the 3D simulation is $\alpha = 0.55$,
roughly consistent with the typical size of the turbulent structures
$l_{\rho}/H_0 \approx 0.3-0.4$ (as assumed in mixing length theory).
The density inversion, present in the 1D radiative equilibrium 
initial condition, is absent in the turbulent state.  This
is because a significant fraction of the luminosity is transported by
convection (Fig. \ref{CompFlux}), which reduces the radiation
acceleration to be sub-Eddington.
For {\sf StarDeep}, the turbulent velocity is very subsonic with the
turbulent kinetic energy less than $1\%$ of the thermal energy
(Fig. \ref{CompEnergy}). The difference between the actual thermal
gradient $\nabla$ and the adiabatic gradient $\nabla_{\rm ad}$ is
smaller than $8\%$ of $\nabla_{\rm ad}$.  All of these properties are
consistent with expectations for efficient convection.  

When $\tau_0<\tau_c$, the advection flux drops significantly as
photons cannot be trapped by the fluid due to rapid diffusion. We see
this transition in the surface layers of run {\sf StarMid}.
As shown in Figure \ref{StarMidConvection}, below $ 60\rsun$, 
the averaged radiation advection flux is consistent with MLT (eq. \ref{eq:mlt}). 
The turbulent velocity is subsonic and $\tilde{\nabla}\ll 1$ as in {\sf StarDeep}. However 
above $ 60 \rsun$, equation~(\ref{eq:mlt}) significantly overpredicts the average convection 
flux if we use a constant $\alpha$, a choice motivated by the fact that the 
correlation length of the turbulence ($l_{\rho}/H_0$) is almost the same through the whole envelope. 
Above  $ 60\rsun$ the turbulent velocity reaches the isothermal sound speed 
and $\tilde{\nabla}$ increases significantly. 

The run {\sf StarTop} has $\tau_0\ll \tau_c$ and convection is very
inefficient.  In this case the average radiation advection flux is
$\ll 1\%$ of the diffusive flux throughout the envelope
(Figure \ref{CompFlux}). If we adopt the non-adiabatic MLT model \citep[e.g.,][]{Henyey:1965,Magicetal2015}
 with $\alpha=0.4$, which is the ratio of the turbulent correlation length to the pressure 
scale height, we find that it still significantly over-predicts the convection flux (Figure \ref{StarTopConvection}). 
The difference is probably caused by a combination of the porosity effect and vertical oscillation 
of the envelope as explained blow. 

The turbulent velocity in {\sf
  StarTop} is larger than the isothermal sound speed and strong shocks
are formed, driving large density fluctuations. The bottom panel of
Figure \ref{CompEnergy} shows, however, that the kinetic energy
density only reaches the gas internal energy density, which is still
much smaller than the radiation energy density.  The large density
fluctuations allow radiation to escape somewhat more efficiently \citep{Shaviv1998}
than through a homogenous medium (``porosity"), reducing the effective
radiation acceleration.  This effect is not, however, large enough to
make the radiation acceleration smaller than the gravitational
acceleration.  The density inversion present in the initial conditions
is initially destroyed by convection and the envelope expands, but the
gas quickly falls back because of the reduced opacity and radiation
acceleration. The density inversion then reforms.  This cycle repeats,
with the envelope undergoing strong vertical oscillations with a
period of a few hours. 
The density inversion predicted by 1D
radiative equilibrium models still exists in the time averaged
vertical profile of {\sf StarTop}, in contrast to {\sf StarDeep} and
{\sf StarMid} where convection is able to largely eliminate the
density inversion.
  
\subsection{Observational consequences}
A signature of inefficient convection is the appearance of supersonic
turbulent velocities with respect to the isothermal sound speed (but
not the radiation sound speed).  This is because rapid diffusion leads
the photons and gas to decouple and the radiation pressure cannot
respond to density fluctuations.  This in turn makes the gas highly
compressible, leading to shocks and large density fluctuations.  



In the calculations that include the photosphere ({\sf StarTop} and
{\sf StarMid}) we observe turbulent velocities reaching the isothermal
sound speed close to the stellar surface ($\approx 50$ km/s).  These
large velocities can impact the spectroscopic measurements of line
profiles, as quantified by the macroturbulence (microturbulence)
parameters. These parameters quantify the velocity fields at the
stellar photosphere with correlation length larger (smaller) than the
line-forming region. The extent of the line forming region is
comparable to the local pressure scale-height, which is about the
correlation length of the turbulent structures in our
calculations. Therefore we expect the velocity fields observed in the
simulations to potentially impact both micro- and macro- turbulence
measurements.  This is important, as the broadening of spectral lines
is used to measure the (projected) rotational velocity of stars, which
in the presence of extra turbulent broadening could be be
systematically higher than the actual rotation rate
\citep[][]{Ramirez-Agudelo2015}.  The idea that inefficient convection
at the iron opacity peak could be responsible for the extra broadening
of spectral lines in hot massive stars has been discussed previously
\citep{Cantiello2009}, where they focused on the role of
gravity waves excited by turbulent convection. 
Recently the
potentially observable signature of turbulent pressure in the inefficient
outer convective regions of massive stars has been highlighted by
\citet{Grassitelli2015}, who showed a correlation between observed macroturbulence
and turbulent pressure (as calculated in the context of 1D MLT).

The vertical oscillations of the envelope in {\sf StarMid} and {\sf StarTop} 
also cause the radiation flux coming from the photosphere to vary with time. 
The standard deviations of the total radiation flux leaving from the photosphere 
are $\approx 13\%$ and $\approx 2.7\%$ 
of the time averaged mean values in {\sf StarMid} and {\sf StarTop} respectively. This means that 
this oscillation can be observed in principle. However, direct comparisons 
with observations require global calculations of the envelope.

Strange mode instabilities are commonly associated with opacity bumps and
density inversions in 1D models of massive stars \citep[e.g.,][]{Kiriakidis1993,Saio2009,Saio2013}.
These standing wave instabilities are fundamentally acoustic in nature, and are generally
associated with a trapping region around the opacity bump/density
inversion.  The excitation mechanism of these modes is due to opacity
variations in the acoustic wave, and is related to the growth mechanism
of traveling acoustic waves discussed by Blaes \& Socrates (2003).
As we briefly noted above in section 4.3.1, conditions in {\sf StarTop} at least
are such that these traveling wave instabilities should be driven, and the
fact that the density inversion persists in the time-averaged structure may
permit the existence of trapped strange modes.  It would be interesting to
do a global stability analysis on the time-averaged structure of {\sf StarMid}
and {\sf StarDeep}, to see if standing acoustic wave instabilities could still
be trapped even in the absence of density inversions.

\subsection{Implications for 1D stellar evolution}
For $\tau_0 \gtrsim \tau_c$, we find that mixing length theory
provides a reasonable description of the convective energy transport
in the 3D simulations.  However, the mixing length parameters we find
($\alpha \approx 0.5$) are noticeably smaller than the canonically adopted
values \citep[$\alpha > 1$, see e.g.][]{Yusof2013,Koheler2015}.   
We also find that as $\tau_0/\tau_c$ decreases, the mixing length parameter
decreases.  This is exactly the {\em opposite} of what has been proposed by
different groups dealing with the numerical difficulties associated
with radiation dominated stellar envelopes in 1D.  They suggested that
convection in this situation should be artificially enhanced
(equivalent to increasing the $\alpha$ parameter with respect to the
canonical values above), justifying this choice with either the need
to remove an ``unphysical density inversion'' \citep{Maeder1987,Ekstrom2012} or the need to
include some (unknown) extra energy transport mechanism
\citep{Paxtonetal2013}. 

Moreover we find that density inversions are present 
 in the time averaged structure of the stellar envelope when $\tau_0\ll \tau_c$,
i.e. when an opacity peak is close to the stellar surface. In the case of the iron opacity peak, 
this occurs for very massive main sequence stars,
though we expect the same physics to occur in cooler objects due to the effect of the H and He opacity bumps (not studied in this work).  
The behavior of these density inversions is quite different as compared to
the steady 1D solution, as they tend to be cyclically destroyed and
reformed and are associated with very large density inhomogeneities.
One important effect that is currently missing in the 1D models is the inclusion  
of a ``porosity factor'', reducing the effective 
radiation acceleration when $\tau_0 < \tau_c$ \citep[e.g.,][]{Paczynski1969}. 
The dependency  of the porosity on optical depth, resolution and magnetic fields will be the focus of future work. 

The turbulent, non-stationary behavior of radiation dominated envelopes
suggests a possible modification of the stellar mass loss relative to
spherically symmetric hydrostatic models \citep[e.g.,][]{Owockietal2004}. 
Unfortunately as we only simulate a local patch of the envelope, we cannot determine
the fate of the mass that leaves our simulation box. The mass loss
rate also cannot be calculated self-consistently based on these
simulations.  This requires global calculations of the envelope to
capture the change of gravitational acceleration and geometric
dilution correctly.

The majority of massive stars are found in binary systems \citep{Sana2012}.
Depending on their orbital separation and the evolution of their radii, these stars can fill their Roche lobe 
and strongly interact already during the main sequence \citep[e.g.,][]{mink2014}. 
As shown in this paper, the energy transport in the radiation dominated envelopes of these stars is not properly modeled in current 1D calculations,
resulting in very uncertain values for the stellar radii. This could have important implications for the evolution 
of massive binaries and, for example,  the predicted rates of BH-BH binary merges, which are promising sources of gravitational waves \citep{Belczynski2014}.

One limitation of the simulations presented here is that we have
neglected the effects of the stellar magnetic field. Recently,
magnetic buoyancy has been found to significantly enhance the
radiation advection flux, even with a relatively weak magnetic field
\citep[][]{Blaesetal2011,Jiangetal2014c}.  A fraction of massive stars
do show significant magnetization \citep[][]{Wadeetal2014}, 
and dynamo generated magnetic fields could be present in the iron convection zone 
\citep[][]{CantielloBraithwaite2011}. Therefore the
effects of magnetic field should be studied in future work.

\section*{Acknowledgements}
Y.F.J thanks Jeremy Goodman, Jim Stone, Shane Davis and all members
of the SPIDER network for helpful discussions.  EQ thanks Mark Krumholz
and Todd Thompson for useful conversations.  This work was supported
by the computational resources provided by the NASA High-End Computing
(HEC) Program through the NASA Advanced Supercomputing (NAS) Division
at Ames Research Center; the Extreme Science and Engineering Discovery
Environment (XSEDE), which is supported by National Science Foundation
grant number ACI-1053575; and the National Energy Research Scientific
Computing Center, a DOE Office of Science User Facility supported by
the Office of Science of the U.S.  Department of Energy under Contract
No. DE-AC02-05CH11231.  Y.F.J. is supported by NASA through Einstein
Postdoctoral Fellowship grant number PF-140109 awarded by the Chandra
X-ray Center, which is operated by the Smithsonian Astrophysical
Observatory for NASA under contract NAS8-03060.  EQ is supported in
part by a Simons Investigator Award from the Simons Foundation.
This project was supported by NASA under TCAN grant number NNX14AB53G
and the NSF under grants PHY 11-25915 and AST 11-09174. 

\bibliography{full_article}

\end{CJK*}

\end{document}